\documentclass[11pt]{article}

\usepackage{amsfonts}
\usepackage{epsfig}
\usepackage{latexsym}
\def\bequ{\begin{equation}}
\def\eequ{\end{equation}}
\def\barr{\begin{array}}
\def\earr{\end{array}}

\def\ben{\begin{equation}}
\def\een{\end{equation}}
\def\bena{\begin{eqnarray}}
\def\eena{\end{eqnarray}}

\def\spa#1{\phantom{\fbox{\rule[-#1cm]{0cm}{0cm}}}}

\def\b1{e^0}

\newcommand{\be}{\begin{equation}}
\newcommand{\ee}{\end{equation}}
\def\bea{\begin{eqnarray}}
\def\eea{\end{eqnarray}}

\def\be{\begin{equation}}
\def\ee{\end{equation}}
\def\bea{\begin{eqnarray}}
\def\eea{\end{eqnarray}}

\def\lesssim{\mathrel{\hbox{\rlap{\hbox{\lower4pt\hbox{$\sim$}}}\hbox{$<$}}}}
\def\gtrsim{\mathrel{\hbox{\rlap{\hbox{\lower4pt\hbox{$\sim$}}}\hbox{$>$}}}}

\begin{document}
\title{{\huge \bf{
A gravito-electromagnetic analogy \\ based on tidal tensors}}}

\author{{L. Filipe Costa and Carlos A. R. Herdeiro}
\footnote{Email addresses: filipezola@fc.up.pt, crherdei@fc.up.pt}\\
\\ {\em Departamento de F\'\i sica e Centro de F\'\i sica do Porto}
\\ {\em Faculdade de Ci\^encias da
Universidade do Porto}
\\ {\em Rua do Campo Alegre, 687,  4169-007 Porto, Portugal}}

\date{May 2007}       %%
\maketitle

%\centerline{ ~hep-th/yymmnnn}

\begin{abstract}
We propose a new approach to a physical analogy between General Relativity and
Electromagnetism, based on \textit{tidal tensors} of both theories.
Using this approach we write a covariant form for the gravitational analogues
of the Maxwell equations. The following realisations of the analogy
are given.
The first
one matches linearised gravitational tidal tensors to exact electromagnetic
tidal tensors in Minkwoski spacetime. The second one matches exact magnetic
gravitational tidal tensors for ultra-stationary metrics to exact magnetic
tidal tensors of electromagnetism in curved spaces. In the third we show that
our approach leads to two-step exact derivation of the Papapetrou force on a
gyroscope. We then establish a new
proof for a class of tensor identities that define invariants of the type
$\vec{E}^2-\vec{B}^2$ and $\vec{E}\cdot\vec{B}$, and we exhibit the invariants
built from tidal tensors in both gravity and electromagnetism. We contrast our
approach with the two gravito-electromagnetic analogies commonly
found in the literature, and argue that it sheds light on the debate about the
limit of validity of one of the analogies, and clarifies issues concerning the
physical interpretation of the other.

\end{abstract}

\newpage

\tableofcontents

\section{Introduction}
Two rather different analogies between classical Electromagnetism and General
Relativity have been presented in the literature, both of which have been
dubbed \textit{gravito-electromagnetism} (see for instance
\cite{Mashhoon:2003ax} and \cite{Maartens:1997fg}).
\begin{itemize}
\item[$\star$] The first one draws an analogy between some components of the
spacetime metric and the electromagnetic potentials, using linearised theory;
\item[$\star$] The second one makes a parallelism between a
decomposition of the
Weyl and the Maxwell tensors in electric and magnetic parts;
\item[$\star$] Besides these two analogies there is a third interesting
connection between relativistic gravity and electromagnetism: the Klein Gordon
equation in ultra-stationary metrics can be mapped to a non-relativistic
Schr\"odinger problem in a time independent magnetic field in a curved space
\cite{thesis,Drukker:2003mg}.
\end{itemize}
The first analogy relies on the obvious similarity between the linearised
Einstein equations, in the harmonic gauge, and Maxwell's equations in the
Lorentz gauge. It is, therefore, presented for perturbations around Minkowski
spacetime and it is clearly not covariant.

The second analogy, on the other hand, relies on two facts. Firstly
that we can
do an irreducible splitting into electric and magnetic parts for both the
Maxwell tensor and the Weyl tensor. Secondly, that we can find a (formally)
similar set of equations for both the electromagnetic parts of the Maxwell
tensor and the electromagnetic parts of the Weyl tensor. Moreover, in analogy
with their electromagnetic counterparts, the electric and magnetic
parts of the
Weyl
tensor define ``invariant quantities'' (in a sense to be precised
below) of the
type $E^2-B^2$ and $E\cdot B$. Thus, this analogy is covariant and it is exact
(i.e
it does not rely on linearised theory).

\subsection{The need for a new, physically transparent approach}
Within each of the approaches presented in literature there are issues
needing a clarification. In the case of the analogy based on linearised
theory,
its limit of validity has been under debate, and there is no consensus about
it. While some authors limit the analogy to stationary configurations
\cite{Harris1991}-\cite{General Relativity}, others argue it can be
extended to
time dependent setups \cite{Mashhoon:2003ax}-\cite{Harmonic Gauge}.
On this last version of the analogy, a set of Maxwell like equations
are derived, which predict the existence of gravitational induction effects
similar to the electromagnetic ones; experiments to detect those
induced fields
have been
proposed \cite{Thorne Braginsky} and, recently, such an experiment has
actually
been performed \cite{ESA}. However, due to the symmetries of gravitational
tidal tensors, as we shall see, such phenomena cannot take place in gravity.

In the case of the second analogy, it is its physical content that is
not clear
in literature. While the electric part of the Weyl tensor (hereafter
denoted by
$\mathcal{E}_{\mu\nu}$) is regarded as a generalisation of the Newtonian tidal
tensor \cite{LozanovskiAarons:99}-\cite{Bertschinger:1994nc}, its magnetic
counterpart (hereafter denoted by $\mathcal{H}_{\mu\nu}$) is not well
understood \cite{LozanovskiAarons:99,Dunsby:1998hd,VandenBergh:2002fb}, but
claimed to be associated with rotation
\cite{Bonnor:95,Lozanovski:99,Maartens:1997fg,Fodor:1998jp,Matarrese:1993zf}
and gravitational radiation
\cite{Matarrese:1992rp}-\cite{Bertschinger:1994nc}.
However, immediately contradictions arise
\cite{Bonnor:95,Lozanovski:99,Maartens:1997fg}: there are many known examples
of rotating spacetimes where the magnetic part of the Weyl tensor vanishes;
amongst them is the notorious
example of the G\"odel Universe. It is also clear that gravitational waves
cannot be the sole source for $\mathcal{H}_{\mu\nu}$, since the latter is
generically non-vanishing in most stationary
spacetimes. Another topic under debate is the Newtonian limit of
$\mathcal{H}_{\mu\nu}$; while some authors assert it must vanish in that limit
(e.g. \cite{Maartens:1998ci,Ellis:Newtonian Weyl}), it has been argued it does
not \cite{Bertschinger:1994nc,Hui:1996}.

The physical connection between these two analogies is another matter needing
for a clarification.  They are not only distinct in their
approach; they seem to refer to different phenomena and (since the physical
content of the second approach is still an open question in literature) one
might even be led to \textit{apparently} contradictory conclusions. That is
what happens,
for instance, in the cases of the the Lense and Thirring or the Heisenberg
spacetimes. According to the first approach, their ``gravito-electric field''
is zero, while their ``gravito-magnetic field'' is finite and uniform. But in
the second approach, these spacetimes are classified as ``purely electric
spacetimes'' since the Weyl tensor is electric.\footnote{The magnetic and
electric parts of the Weyl tensor are observer dependent. Both in the
Lense and
Thirring spacetime and the Heisenberg group manifold there are observers for
which the magnetic part of the Weyl tensor vanishes, while the electric part
never vanishes. For this reason, these spacetimes are classified as ``purely
electric'' (see for instance  \cite{Bonnor:1995zf}). The electric character of
the spacetime is equivalently revealed by the invariants, which using the
notation of section
  \ref{secanalogy2} are $\mathcal{ L}>0$, $\mathcal{M}=0$.}

\subsection{A new approach to gravito-electromagnetism based on tidal tensors}
The purpose of this paper is to propose a new approach to
gravito-electromagnetism. We claim that a \textit{physical analogy}
steams from
the tidal tensors of the two theories. This analogy stands on universal,
covariant equations: the geodesic deviation equation and its analogue
electromagnetic worldline deviation equation; the Papapetrou force
applied on a
gyroscope and the electromagnetic force exerted on a magnetic dipole. The
identification of the correct tidal tensors allows us to write an explicitly
covariant form for the Maxwell equations, and derive their gravitational
analogues. From these tidal tensors we also define, whenever they exist,
analogous electromagnetic and gravitational invariants.
Explicit evidence for this analogy is additionally given by a
comparison of the
electromagnetic and gravitational tidal tensors, as well as the
invariants from
them constructed, in an elementary example of analogous physical systems (a
spinning mass in gravity, and a spinning charge in electromagnetism); in
generic electromagnetic fields and linearised gravitational perturbations; and
by a suggestive application of the Klein-Gordon equation in ultra-stationary
spacetimes.

The approach proposed herein clarifies several issues concerning the
gravito-electromagnetic approaches found in the literature:
\begin{itemize}
\item[$\star$] While embodying all the correct predictions from the
usual linear
perturbation approach, our approach reveals, in an unambiguous way, its regime
of validity (for which, as mentioned before, there is no consensus in the
literature);
\item[$\star$] It sheds light into the debate about the physical
content of the
second approach. In analogy with the electromagnetic tidal tensors, we give a
simple physical interpretation for the electric and magnetic parts of the Weyl
tensor, which trivially solves the contradictions discussed in the literature;
\item[$\star$] Finally, the third connection mentioned above becomes
but another
realisation of this analogy, even somewhat more surprising because there is an
exact matching between (magnetic) tidal tensors of a non-linear theory
(gravity)
and
the ones of a linear theory (electromagnetism).
\end{itemize}

This paper is organised as follows. Our proposal, of a \textit{physical
analogy} between gravity and electromagnetism based on tidal tensors is
presented in section 2, where a covariant form for the gravitational analogues
of the Maxwell equations is derived and its physical implications discussed.
Three realisations
follow. In the first we study the analogy for small perturbations around
Minkowski spacetime, and compare gravitational and electromagnetic tidal
tensors in an example of physically analogous setups. The second gives an
interpretation for the physical similarities of the
Klein-Gordon in ultra-stationary spacetimes with the Schr\"odinger equation in
some curved spaces. Finally, in the third we show that within the framework of
our approach, the exact Papapetrou equation for the force applied on gyroscope
is obtained in a ``two step'' derivation, avoiding the lengthy original
computation, and gaining physical insight.  In sections \ref{linearth} and
\ref{secanalogy2} the aforementioned two approaches to
gravito-electromagnetism
are reviewed and contrasted with our proposal. The second analogy is used to
introduce gravitational
invariants of the type of $\vec{E}^2-\vec{B}^2$ and $\vec{E}\cdot\vec{B}$,
which are `deconstructed'  in section \ref{secinv}, where the reason for such
scalar
invariants to be observer independent is dissected and new invariants,
the ones
that play a similar role in the gravito-electromagnetic analogy based on tidal
tensors are constructed. We close with a discussion.

\section{The physical gravito-electromagnetic analogy, based on tidal tensors}
\label{the analogy based on tidal tensors}
An analogy with physical content between electromagnetism and general
relativity
must, if it exists, be based on physical quantities common to both theories.
Taking the perspective that, in general relativity, the only `physical forces'
(in the sense of being covariant) are the tidal forces, described by the
curvature tensor, the starting point of the analogy should be the tidal
tensors. In this section we build an analogy between the gravitational and
electromagnetic tidal tensors. For this purpose we start by defining these
tensors.

\subsection{Gravitational and Electromagnetic Tidal Tensors}
\label{Gravitational and Electromagnetic Tidal Tensors}

Tidal forces in gravity are described in a covariant way by the \emph{geodesic
deviation} equation:\begin{equation}
\frac{D^{2}\delta
x^{\alpha}}{D\tau^{2}}=-R_{\,\,\,\mu\beta\sigma}^{\alpha}U^{\mu}U^{\sigma}\delta
x^{\beta}\label{desvio geo}\end{equation}
where $D/D\tau$ denotes covariant differentiation along a curve parameterised
by $\tau$. This equation gives the relative acceleration of two neighbouring
particles with the same 4-velocity $U^{\alpha}$ (see Appendix
\ref{Worldlinedev}). In
order to find the electromagnetic analogue to (\ref{desvio geo}),
one must first notice a very intrinsic difference between the two
interactions: while the ratio between gravitational and inertial mass
is universal, the same does not apply to the ratio between electrical
charge and inertial mass; i.e., there is no electromagnetic counterpart
of the equivalence principle. Therefore, the analogue electromagnetic
problem will be to consider two neighbouring particles with the same
4-velocity $U^{\alpha}$ in an electromagnetic field on Minkowski
spacetime, with the extra condition that the two particles have the
same $q/m$ ratio. Under these conditions one obtains the \emph{worldline
deviation} equation (see Appendix \ref{Worldlinedev}):\begin{equation}
\frac{D^{2}\delta
x^{\alpha}}{D\tau^{2}}=\frac{q}{m}F_{\,\,\,\mu;\beta}^{\alpha}U^{\mu}\delta
x^{\beta}\label{desvio EM}\end{equation}
where $F_{\alpha\beta}$ is the Maxwell tensor. The comparison
of (\ref{desvio geo}) and (\ref{desvio EM}) suggests a physical
analogy between the two rank-2 tensors:\begin{equation}
\mathbb{E}_{\alpha\beta}\equiv
R_{\alpha\mu\beta\sigma}U^{\mu}U^{\sigma}\longleftrightarrow
E_{\alpha\beta}\equiv F_{\alpha\mu;\beta}U^{\mu}\label{Eab}\end{equation}

The tensor $E_{\alpha\beta}$ is the covariant derivative of the electric field
$E^{\alpha}=F^{\alpha\beta}U_{\beta}$ seen by the observer of (fixed)
4-velocity field $U^{\alpha}$;
for this reason we will refer to it as the \emph{electric tidal tensor}
and its gravitational counterpart $\mathbb{E}_{\alpha\beta}$, which
is known in literature as \emph{electric part of the Riemann tensor}
(cf. section 4), as the \emph{electric gravitational tidal tensor.}
The different signs in (\ref{desvio geo}) and (\ref{desvio EM})
reflect the different character (attractive or repulsive) of the interaction
between masses or charges of the same sign. Given our definition of
the electric tidal tensor, it is straightforward to define the \emph{magnetic
tidal tensor}\begin{equation}
B_{\alpha\beta}\equiv\star
F_{\alpha\mu;\beta}U^{\mu}=\frac{1}{2}\epsilon_{\:\,\,\,\,\alpha\mu}^{\gamma\lambda}F_{\gamma\lambda;\beta}U^{\mu}\label{Bab}\end{equation}
where $\star$ denotes the Hodge dual and $\epsilon_{\alpha\beta\gamma\delta}$
is the Levi-Civita tensor. This tensor measures the tidal effects produced by
the magnetic field $B^{\alpha}=\star F^{\alpha\beta}U_{\beta}$ seen by the
observer of 4-velocity $U^{\gamma}$.
An analogous procedure applied to the Riemann tensor yields the so
called \emph{magnetic part of the Riemann tensor} (cf. section
4)\begin{equation}
\mathbb{H}_{\alpha\beta}\equiv\star
R_{\alpha\mu\beta\sigma}U^{\mu}U^{\sigma}=\frac{1}{2}\epsilon_{\:\,\,\,\,\alpha\mu}^{\gamma\lambda}R_{\gamma\lambda\beta\sigma}U^{\mu}U^{\sigma}\label{Hab}\end{equation}
which we claim, and give evidence throughout this paper, to be the
\emph{physical} gravitational analogue of $B_{\alpha\beta}$:\[
B_{\alpha\beta}\longleftrightarrow\mathbb{H}_{\alpha\beta}\]
For this reason $\mathbb{H}_{\alpha\beta}$ will be herein referred
to as the \emph{magnetic gravitational tidal tensor}. In (\ref{Hab})
the Hodge dual was taken with respect to the first pair of indices
of the Riemann tensor; a different choice amounts to changing the
order of the indices in $\mathbb{H}_{\alpha\beta}$. Note that this
tensor is generically \textit{not} symmetric, only in vacuum.

\subsubsection{Maxwell equations as tidal tensor equations}
\label{Maxwell equations as tidal tensor equations}

Maxwell equations are tidal equations; indeed, an explicitly covariant
form for them can be obtained in a simple fashion from the above defined
electromagnetic tidal tensors\footnote{Here and throughout the paper
(except in
sec. \ref{linearth}) we use $c=1=G$.}:\begin{equation}
\begin{array}{lll}
\mbox{i)}\, E_{\,\,\,\alpha}^{\alpha}=4\pi\rho_{c} &  & \mbox{ii)}\,
E_{[\alpha\beta]}=\frac{1}{2}F_{\alpha\beta;\gamma}U^{\gamma}\\
\\ \mbox{iii)}\, B_{\,\,\,\alpha}^{\alpha}=0 &  & \mbox{iv)}\,
B_{[\alpha\beta]}=\frac{1}{2}\star
F_{\alpha\beta;\gamma}U^{\gamma}-2\pi\epsilon_{\alpha\beta\sigma\gamma}j^{\sigma}U^{\gamma}\end{array}\label{MaxCov}\end{equation}
where $j^{\alpha}$ and $\rho_{c}=-j^{\alpha}U_{\alpha}$ denote,
respectively, the (charge) current 4-vector and the charge density
as measured by the observer of 4-velocity $U^{\alpha}$. These equations can be
completely expressed in terms of tidal tensors and sources, by noting that:
\begin{eqnarray}
F_{\alpha\beta;\gamma}U^{\gamma} & = &
2U_{[\alpha}E_{\beta]\gamma}U^{\gamma}+\epsilon_{\alpha\beta\mu\sigma}U^{\sigma}B^{\mu\gamma}U_{\gamma}\nonumber\\
\star F_{\alpha\beta;\gamma}U^{\gamma} & = &
2U_{[\alpha}B_{\beta]\gamma}U^{\gamma}-\epsilon_{\alpha\beta\mu\sigma}U^{\sigma}E^{\mu\gamma}U_{\gamma}\label{Fcov
Fstarcov}\end{eqnarray}
which follows from decomposition
(\ref{decompF}) given in section 4 by a straightforward computation.
Equations (\ref{MaxCov}i) and (\ref{MaxCov}iii) are the covariant forms of
$\nabla.\vec{E}=\rho_{c}$ and $\nabla.\vec{B}=0$; equations (\ref{MaxCov}ii)
and (\ref{MaxCov}iv) are covariant forms for
$\nabla\times\vec{E}=-\partial\vec{B}/\partial t$
and $\nabla\times\vec{B}=\partial\vec{E}/\partial t+4\pi\vec{j}$,
respectively.

\subsubsection{The Gravitational Analogue of Maxwell equations}
\label{The Gravitational Analogue of Maxwell equations}

In what follows it proves useful to introduce the decomposition of
the Riemann tensor (eg. \cite{cherubini:02}):\begin{equation}
R_{\alpha\beta\gamma\delta}=C_{\alpha\beta\gamma\delta}+g_{\alpha[\gamma}R_{\delta]\beta}+g_{\beta[\delta}R_{\gamma]\alpha}+\frac{1}{3}g_{\alpha[\delta}g_{\gamma]\beta}R\label{Riemann-Weyl}\end{equation}
where $C_{\alpha\beta\gamma\delta}$ is the Weyl tensor, which (by
definition) is traceless and exhibits the property: $\star
C_{\alpha\beta\gamma\delta}=C\star_{\alpha\beta\gamma\delta}$.
$\mathbb{E}_{\alpha\beta}$ and $\mathbb{H}_{\alpha\beta}$ then
become:\begin{equation}
\mathbb{E}_{\alpha\beta}=\mathcal{E}_{\alpha\beta}+\frac{1}{2}\left[g_{\alpha\beta}R_{\gamma\delta}U^{\gamma}U^{\delta}-R_{\alpha\beta}-2U_{(\alpha}R_{\beta)\delta}U^{\delta}\right]+\frac{R}{6}\left[g_{\alpha\beta}+U_{\alpha}U_{\beta}\right]\label{ERiemann-EWeyl}\end{equation}
\begin{equation}
\mathbb{H}_{\alpha\beta}=\mathcal{H}_{\alpha\beta}+\frac{1}{2}\epsilon_{\alpha\beta\sigma\gamma}R_{\textrm{\,\,}\delta}^{\sigma}U^{\gamma}U^{\delta}\label{HRiemann-HWeyl}\end{equation}
where
\bequ
\mathcal{E}_{\alpha\beta}\equiv
C_{\alpha\gamma\beta\sigma}U^{\gamma}U_{\textrm{
}}^{\sigma}, \ \ \ \ \ \mathcal{H}_{\alpha\beta}\equiv\star
C_{\alpha\gamma\beta\sigma}U^{\gamma}U_{\textrm{ }}^{\sigma} ,
\label{Weyl E H}
\eequ
are, respectively, the electric and magnetic parts of the Weyl tensor, both of
which
are symmetric and traceless.

By repeating the same procedure that led to equations (\ref{MaxCov}),
i.e, by taking the traces and anti-symmetric parts of the tidal tensors,
we obtain the analogue set of equations:\begin{equation}
\begin{array}{lll}
\mbox{i)}\,\mathbb{E}_{\,\,\,\alpha}^{\alpha}=4\pi\left(2\rho_{m}+T_{\,\,\alpha}^{\alpha}\right)
&  & \mbox{ii)}\,\mathbb{E}_{[\alpha\beta]}=0\\
\\ \mbox{iii)}\,\mathbb{H}_{\,\,\,\alpha}^{\alpha}=0 &  &
\mbox{iv)}\,\mathbb{H}_{[\alpha\beta]}=-4\pi\epsilon_{\alpha\beta\sigma\gamma}J^{\sigma}U^{\gamma}\end{array}\label{MaxGrav}\end{equation}
where $T_{\alpha\beta}$ denotes the energy momentum tensor, and
$J^{\alpha}=-T_{\
\beta}^{\alpha}U^{\beta}$ and
$\rho_{m}=T_{\alpha\beta}U^{\alpha}U^{\beta}$ are,
respectively,
the mass/energy current density and the mass/energy density as measured
by the observer of four velocity $U^{\alpha}$.

\subsubsection{Gravity versus Electromagnetism}
\label{Gravity vs Electromagnetism}

Equations (\ref{MaxCov}) are strikingly similar to equations (\ref{MaxGrav})
when the setups are stationary in the observer's rest frame. Otherwise,
comparing the two sets of equations tells us that gravitational and
electromagnetic interactions must differ significantly, since the
tidal tensors do not have the same symmetries. We shall now discuss
each equation in detail.

Comparing (\ref{MaxCov}i) and (\ref{MaxGrav}i), we see that the
gravitational analogue of the electric charge density $\rho_{c}$
is $2\rho_{m}+T_{\,\,\alpha}^{\alpha}$. Thus, changing the sign of
the latter combination amounts to changing the character of the gravitational
source from attractive (positive) to repulsive (negative); requiring
it to be positive is the statement of the strong energy condition.
The analogy gets more enlightening if we use the energy momentum tensor
of a perfect fluid (energy density $\rho_{m}$, pressure $p$); in
that case we have the correspondence $\rho_{c}\leftrightarrow\rho_{m}+3p$,
manifesting the contribution of pressure as a source of the gravitational
field. The combination $\rho_{m}+3p$ is well known from the Raychaudhuri
equation of FRW models, determining if the expansion of the universe
is decelerated or accelerated.

A perfect analogy exists in the case of right eqs. of (\ref{MaxCov}iii)
and (\ref{MaxGrav}iii): the trace of $B_{\alpha\beta}$ is zero by
virtue of the electromagnetic Bianchi identity; likewise, the trace
of $\mathbb{H}_{\alpha\beta}$ vanishes by virtue of the first Bianchi
identities.

Equations (\ref{MaxCov}ii) and (\ref{MaxGrav}ii) reveal a fundamental
difference between $\mathbb{E}_{\mu\gamma}$ and $E_{\mu\gamma}$:
while the former is always symmetric, the latter is only symmetric
if the Maxwell tensor is covariantly constant along the observer's
worldline. The physical content of these equations depends crucially
on these symmetries: since (\ref{MaxCov}ii) is a covariant form of
$\nabla\times\vec{E}=-\partial\vec{B}/\partial t$, the statement
encoded in the equation $\mathbb{E}_{[\alpha\beta]}=0$ is that there
is no gravitational analogue to Faraday's law of induction.

There is a clear gravitational counterpart to the Amp\'ere law: in stationary
(in the observer rest frame) configurations, equations (\ref{MaxCov}iv)
and (\ref{MaxGrav}iv) match up to a factor of 2; therefore, currents
of mass/energy source gravitomagnetism just like currents of charge source
magnetism.
The extra factor of 2 in (\ref{MaxGrav}iv) reflects the different
spin of the gravitational and electromagnetic interactions; in some literature
the gravito-magnetic charge $Q_B=2m$ (e.g. \cite{larmor,Mashhoon:2003ax}) has
been defined to account for this factor.
In generic dynamics, again, the physical content of these equations will be
drastically
different, due to the presence of the induction term $\star
F_{\alpha\beta;\gamma}U^{\gamma}$
in (\ref{MaxCov}iv), which has no counterpart in (\ref{MaxGrav}iv).
This means that there is no gravitational analogue to the magnetic fields
induced, for instance, by the time varying electric field between the
plates of
a charging/discharging
capacitor.

The different symmetries of the gravitational and electromagnetic
tidal tensors are related to a fundamental difference in their tensorial
structure: while the former are \textit{spatial}, the latter are \textit{not}. The first
statement means that if, say, $\mathbb{A}_{\alpha\beta}$ is a gravitational
tidal tensor as measured by the observer of four velocity $U^{\alpha}$,
then $\mathbb{A}_{\alpha\beta}U^{\alpha}=\mathbb{A}_{\alpha\beta}U^{\beta}=0$,
which trivially follows from the symmetries of the Riemann tensor.
In electromagnetism, we have a different situation. On one hand,
$E_{\alpha\beta}U^{\alpha}=0$,
which is trivial consequence of the symmetries of $F_{\alpha\beta}$;
but, on the other hand, $E_{\alpha\beta}U^{\beta}$ is non-zero when
the field varies with the observer proper time. The same applies to
$B_{\alpha\beta}$. These contractions determine the temporal projections
of the electromagnetic tidal tensors, and they are indeed at
the origin of the extra terms in (\ref{MaxCov}iv) and (\ref{MaxCov}ii)
as compared to (\ref{MaxGrav}ii) and (\ref{MaxGrav}iv), as can be seen from
expressions (\ref{Fcov Fstarcov}).

\subsection{Small perturbations around Minkowski spacetime}
\label{minper}
Let us start by considering a generic electromagnetic field, described by the
potential one-form
\bequ
A=-\phi(t,x^i)dt + A_j(t,x^i)dx^j \ , \label{pot1}\eequ
in Minkowski space, with metric
\bequ
ds^2=-dt^2+\hat{g}_{ij}(x^k)dx^idx^j \ , \label{Minkowski} \eequ
where $\hat{g}_{ij}$ is an arbitrary spatial metric on $\mathbb{R}^3$. It
follows that
\[ F=dA=(\phi_{;i}+\dot{A}_i)dt\wedge dx^i+A_{j;i}dx^i\wedge dx^j \ ,
\]
\[\star F=A^{j;i}\hat{\epsilon}_{ijk}dx^k\wedge
dt+\frac{1}{2}\hat{\epsilon}_{ijk}(\phi^{;i}+\dot{A}^i)dx^j\wedge dx^k \ ,
\]
where dots represent time derivatives, the semi-colon represents covariant
derivatives with respect to $\hat{g}_{ij}$ (covariant derivatives in
$\hat{g}_{ij}$ commute since the metric is flat) and
$\hat{\epsilon}_{ijk}$ are
the components of the Levi-Civita tensor on $\mathbb{R}^3$ in coordinates
$\{x^i\}$. We take the orientation defined by $\epsilon_{0123}=-1$. The
electric tidal tensor $E_{\alpha\beta}$ (\ref{Eab})
is, for an observer with four velocity $U^{\alpha}=(u^0,u^i)$, given by
\bequ
\barr{c}
\displaystyle{ E_{00}=(\dot{\phi}_{;i}+\ddot{A}_i)u^i \ , \ \ \ \ \ \ \ \ \
E_{ij}=-(\phi_{;ij}+\dot{A}_{i;j})u^0+2A_{[k;i]j}u^k \ ,} \spa{0.4cm}\\
\displaystyle{E_{i0}=-(\dot{\phi}_{;i}+\ddot{A}_i)u^0+2\dot{A}_{[j;i]}u^j \ ,
\
\ \ \ \ E_{0i}=(\phi_{;ki}+\dot{A}_{k;i})u^k \ . } \earr
\label{Egeneral} \eequ
Similarly, the magnetic tidal tensor (\ref{Bab}) is given by
\bequ
\barr{c}
\displaystyle{ B_{00}=-\hat{\epsilon}_{ijk}\dot{A}^{j;i} u^k \ , \ \ \
\ \ \ \ \
\ B_{ij}=\hat{\epsilon}_{lmi}A^{m;l}_{\ \ \ ;j}
u^0+\hat{\epsilon}_{lik}\left(\phi^{;l}_{\ ;j}+\dot{A}^l_{\ ;j}\right)u^k \ ,}
\spa{0.4cm}\\ \displaystyle{B_{i0}=\hat{\epsilon}_{kji}\dot{A}^{j;k}
u^0+\hat{\epsilon}_{jik}\left(\dot{\phi}^{;j}+\ddot{A}^j\right)u^k \ , \ \ \ \
\ B_{0i}=-\hat{\epsilon}_{ljk}A^{j;l}_{\ \  ;i} u^k \ . } \earr
\label{Bgeneral}
\eequ
Obviously, these quantities would look simpler and more familiar if written in
terms of the electric and magnetic fields. But writing them in this fashion
makes the comparison with the gravitational case explicit.

Now consider linearised perturbations of Minkowski spacetime
(\ref{Minkowski}).
We take the perturbed metric in the form\footnote{Here we have re-inserted the
velocity of light `c', despite the fact that in this subsection we are putting
$c=1$. The $c$ dependence will, however, be of use in section 3.}
\bequ
ds^2=-c^2(1-2\frac{\phi(t,x^i)}{c^2})dt^2-\frac{4}{c}A_j(t,x^i)dtdx^j+\left[\left\{1+2\frac{C(t,x^k)}{c^2}\right\}\hat{g}_{ij}(x^k)+2\frac{\xi_{ij}(t,x^k)}{c^2}\right]dx^idx^j
\ , \label{mper}\eequ
where, as before, $\hat{g}_{ij}$ is an arbitrary spatial metric on
$\mathbb{R}^3$ and $\xi_{ij}$ is traceless. A simple computation reveals that (again, the semi-colon represents covariant
derivatives with respect to $\hat{g}_{ij}$):
\[ R^{0}_{\  i 0j}=\ddot{C}\hat{g}_{ij}+\phi_{;ji}+2\dot{A}_{(j;i)}+
\ddot{\xi}_{ij} \ , \ \ \ \ R^{0}_{\
ijk}=-2\hat{g}_{i[j}\dot{C}_{;k]}+2A_{[k;j]i}+2\dot{\xi}_{i[k;j]} \ , \]
\[  R^i_{\  jkl}=-2C_{;j[l}\delta^i_{k]}+\hat{g}_{j[k}C^{;i}_{\
;l]}+2\xi^{i}_{\
[l;k]j}-2\xi_{j[l;k]}^{\ \ \ \ ;i} \ . \]
The perturbations actually separate into scalar, vector and tensor parts
\cite{stewart}, but such separation will not be needed here. The dual Riemann
tensor components are:
\[ \star
R_{0l0i}=-\hat{\epsilon}_{lik}\dot{C}^{;k}+\hat{\epsilon}_{ljk}\left(A^{k;j}_{\
\ ;i}+\dot{\xi}^{\ k;j}_{i}\right) \ ,\ \ \ \  \star
R_{ijk0}=\hat{\epsilon}_{ij}^{\ \
l}\left(\ddot{C}\hat{g}_{lk}+\phi_{;kl}+2\dot{A}_{(k;l)}+
\ddot{\xi}_{lk}\right)
\ , \]
\[ \star R_{k0ij}=\hat{\epsilon}_{kl}^{\ \ m}\left(\hat{g}_{m[i}C^{;l}_{\
;j]}-C_{;m[j}\delta^{l}_{i]}+\xi_{[j;i]}^{l \ \  ;m}-\xi_{m[j;i]}^{\ \ \ \ \
;l}\right) \ , \
\star R_{ijkl}=2\hat{\epsilon}_{ij}^{\ \
m}\left(\hat{g}_{m[k}\dot{C}_{;l]}-A_{[l;k]m}-\dot{\xi}_{m[l;k]}\right) . \]
The electric gravitational tidal tensor $\mathbb{E}_{\alpha\beta}$ (\ref{Eab})
is, for an observer with four velocity $U^{\alpha}=(u^0,u^i)$, given by
\bequ
\barr{c}
\displaystyle{\mathbb{E}_{00}=-\left(\ddot{C}\hat{g}_{ij}+\phi_{;ji}+2\dot{A}_{j;i}+
\ddot{\xi}_{ij}\right)u^iu^j \ ,} \spa{0.5cm}\\
\displaystyle{\mathbb{E}_{i0}=\mathbb{E}_{0i}=\left(\ddot{C}\hat{g}_{ij}+\phi_{;ij}+2\dot{A}_{(i;j)}+
\ddot{\xi}_{ij}\right)u^0u^j+2\left(\hat{g}_{j[i}\dot{C}_{;
k]}-A_{[k;i]j}-\dot{\xi}_{j[k;i]}\right)u^ku^j \ ,}
\earr
\label{E00 E0i}
\eequ
\bequ
\barr{l}
\displaystyle{
\mathbb{E}_{ij}=-(\ddot{C}\hat{g}_{ij}+\phi_{;ij}+2\dot{A}_{(i;j)}+
\ddot{\xi}_{ij})(u^0)^2} \spa{0.3cm}\\
\displaystyle{~~~~~~~+2(\hat{g}_{k(i}\dot{C}_{;j)}-\hat{g}_{ij}\dot{C}_{;k}+A_{k;ij}-A_{(i;j)k}+\dot{\xi}_{k(i;j)}-\dot{\xi}_{ij;k})u^0u^k}\spa{0.3cm}\\
\displaystyle{~~~~~~~
+2u^ku_{(i}C_{;j)k}-(\hat{g}_{ij}C_{;lk}+\hat{g}_{kl}C_{;ij}-2\xi_{l(i;j)k}+\xi_{ij;lk}+\xi_{lk;ij})u^ku^l
\ . }
\earr
\label{Eij}
\eequ
The magnetic gravitational tidal tensor (\ref{Hab}) is,
\begin{equation}
\mathbb{H}_{00}=\hat{\epsilon}_{imn}\left(A^{n;m}_{\ \ \ ;j}+\dot{\xi}^{\
n;m}_{j}\right)u^iu^j \ , \ \ \ \
\mathbb{H}_{ij}=\left(-\hat{\epsilon}_{ijk}\dot{C}^{;k}+\hat{\epsilon}_{ilk}\left(A^{k;l}_{\
\ ;j}+\dot{\xi}^{\ k;l}_{j}\right)\right)(u^0)^2 \label{H00 Hij}
\end{equation}
\bequ
\barr{l}
\displaystyle{
\mathbb{H}_{i0}=\left(\hat{\epsilon}_{ijk}\dot{C}^{;k}-\hat{\epsilon}_{ilk}\left(A^{k;l}_{\
\ ;j}-\dot{\xi}^{\ k;l}_{j}\right)\right)u^0u^j -\hat{\epsilon}_{ik}^{\ \
l}\left(\ddot{C}\hat{g}_{lj}+\phi_{;jl}+2\dot{A}_{(l;j)}+
\ddot{\xi}_{lj}\right)u^ku^j \,}
\earr
\eequ
\bequ
\barr{l}
\displaystyle{
\mathbb{H}_{0i}=\left(\hat{\epsilon}_{jik}\dot{C}^{;k}-\hat{\epsilon}_{jlk}\left(A^{k;l}_{\
\ ;i}-\dot{\xi}^{\ k;l}_{i}\right)\right)u^0u^j } \spa{0.3cm}\\
\displaystyle{~~~~~~~-\hat{\epsilon}_{jl}^{\ \ m}\left(\hat{g}_{m[i}C_{;k]}^{\
;l}+\delta^{l}_{[k}C_{;i]m}+\xi_{\ [k;i]m}^{l}+\xi_{m[i;k]}^{\ \ \ \ \
;l}\right)u^ju^k \ .}
\earr
\label{H0i}
\eequ

As can be now easily verified, the electromagnetic tidal tensors are,
generically, very different from their gravitational counterparts. But if one
takes time independent electromagnetic potentials/gravitational perturbations,
and a "static observer"\footnote{We define the ``static observer'' as being an
observer for which the setup is stationary. An observer at rest
relative to the
centre of mass of the spinning spheres considered in \ref{rotatingp} is an
example of such an observer.}, $U^{\mu}=\delta^{\mu}_0$, then
\bequ
E_{ij}=-\phi_{;ij}=\mathbb{E}_{ij}  \ , \ \ \ \ \
B_{ij}=\hat{\epsilon}_{lki}A^{k;l}_{\ \ ;j}=\mathbb{H}_{ij} \ .
\label{mgtt}\eequ

One may regard this matching between the tidal tensors of the two
theories as an
analogy between the electromagnetic potential $A_{\mu}$ and some components of
the metric tensor, and hence define the ``gravito-electromagnetic fields'':
\begin{equation}
E_G^{i}=-\phi^{;i} \ , \ \ \ \ \ B_{G}^{i}=\hat{\epsilon}^{lki}A_{k;l}
\label{E B nossa analogia} \eequ
These fields help us visualise geodesic motion and frame dragging in analogy
with the more familiar picture of a charged particle subject to a
electromagnetic Lorentz force. Indeed, the geodesic equation
$DU^{\alpha}/D\tau=0$ yields, to linear order in the velocity of the test
particle, the space components:
\begin{equation}
\frac{d^{2}\vec{x}}{dt^{2}}=-\vec{E}_{G}-2\vec{v}\times\vec{B}_{G}\label{geolorentz}
\end{equation}
The ``gravito-magnetic'' field in (\ref{E B nossa analogia}) also leads
directly
to the Lense
and Thirring precession for test gyroscopes \cite{lense,Gravitation and
Spacetime,Gravitation and Inertia}:
as for a magnetic dipole placed in a magnetic field, a {}``torque''
$\tau=-\vec{S}\times\vec{B}_{G}$ acts on a gyroscope of angular
momentum $\vec{S}$ (since in gravity $\vec{S}$ plays the role of the magnetic
moment; see discussion in section \ref{rotatingp} below), causing it to
precess
with angular frequency
$\vec{\omega}=\vec{B}_{G}$, which is accurate to linear order.

We must however stress the fact that this construction is restricted to a
mapping of static electromagnetic fields to stationary gravitational setups,
and is clearly non-covariant: as readily seen by comparing equations
(\ref{Egeneral})-(\ref{Bgeneral}) with (\ref{E00 E0i})-(\ref{H0i}), tidal
tensors will not match for any observer other than the above considered static
observer  $U^{\mu}=\delta^{\mu}_0$.

\subsubsection{Spinning charge versus spinning mass}
\label{rotatingp}
Let us choose an elementary example of analogue physical systems, namely a
rotating spherical mass in General Relativity and a rotating spherical
charge in
Electromagnetism.

Consider a sphere of charge $q$, mass $m$, rotating with constant angular
momentum $J{\bf e}_z$ in Minkowski spacetime. Taking the leading electric
monopole and magnetic dipole contributions, the potential one form becomes:
\[
A=-\frac{q}{r}dt+\frac{\mu \sin^2\theta}{r}d\phi \ , \]
where $\mu\equiv Jq/(2m)$ is the magnetic dipole moment of the rotating
sphere.
Then the Maxwell tensor and its dual are:
\[F=-\frac{q}{r^2}dt\wedge dr-\frac{\mu}{r^2}\sin^2\theta dr\wedge
d\phi+\frac{\mu}{r}\sin 2\theta d\theta\wedge d\phi \ , \]

\[ \ \star F=-q\sin\theta d\theta\wedge d\phi-\frac{\mu}{r^2}dt\wedge
\left(\sin\theta d\theta +\frac{2\cos\theta}{r}dr\right) \ .  \]
For the static observer with four velocity $U^{\mu}=\delta^{\mu}_0$, the
electric and magnetic
tidal tensors are symmetric with
\bequ
\barr{c}
\displaystyle{E_{\alpha
\beta}dx^{\alpha}dx^{\beta}=-\frac{2q}{r^3}dr^2+\frac{q}{r}d\Omega_2} \ ,
\spa{0.4cm}\\ \ \ \ \displaystyle{ B_{\alpha
\beta}dx^{\alpha}dx^{\beta}=\frac{3\mu}{r^2}\left(-\frac{2\cos\theta}{r^2}dr^2+\cos\theta
d\Omega_2-\frac{2\sin\theta}{r}drd\theta\right)} \ .  \earr
\label{sphere} \eequ

Now consider the metric outside a rotating spherical mass, which is
asymptotically described by the Kerr solution \cite{Kerr}. In
Boyer-Linquist coordinates the line element takes the form
\[
ds^2=-\frac{\Delta}{\Sigma}\left(dt-a\sin^2\theta
d\phi\right)^2+\frac{\sin^2\theta}{\Sigma}\left(adt-(r^2+a^2)d\phi\right)^2+\frac{\Sigma}{\Delta}dr^2+\Sigma
d\theta^2 \ ,  \]
where
\[\Delta\equiv r^2-2mr+a^2 \ , \ \ \Sigma=r^2+a^2\cos^2\theta \ . \]
This metric is asymptotically flat. Thus it allows us to compare its
asymptotic
gravitational tidal tensors with the previous electromagnetic tidal tensors.
For a static observer, with four velocity $U^{\mu}\simeq\delta^{\mu}_0$, the
electric and magnetic gravitational tidal tensors are, asymptotically
\bequ
\barr{c}
\displaystyle{\mathbb{E}_{\alpha \beta}dx^{\alpha}dx^{\beta}\simeq
-\frac{2m}{r^3}dr^2+\frac{m}{r}d\Omega_2+\mathcal{O}\left(\frac{1}{r^4}\right)drd\theta}
\ , \spa{0.4cm}\\ \displaystyle{
\mathbb{H}_{\alpha \beta}dx^{\alpha}dx^{\beta}\simeq
\frac{3J}{r^2}\left(-\frac{2\cos\theta}{r^2}dr^2+\cos\theta
d\Omega_2-\frac{2\sin\theta}{r}drd\theta\right)} \ , \earr \label{kerr} \eequ
where $J=ma$ is the physical angular momentum of the spacetime. The electric
gravitational and electromagnetic tidal tensors (\ref{kerr}) and
(\ref{sphere}) match approximately identifying $q\leftrightarrow m$; the
magnetic
ones match up to a factor of 2; this factor, already manifest in
(\ref{MaxGrav}iv), is related to the fact that gravity is a spin 2
interaction,
while electromagnetism is spin 1. It can be interpreted by drawing the analogy
$\mu\leftrightarrow J$ (i.e., assigning to gravity a gyromagnetic ratio equal
to 1).

Now let's see what happens when the observer moves relative to the
spheres of mass/charge; for simplicity we will consider an observer
moving radially with 3-velocity $\vec{v}=v\vec{e}_{r}$, such that
its four velocity is, in the Minkowski spacetime
$U^{\alpha}=\gamma\left(\delta_{0}^{\alpha}+v\delta_{r}^{\alpha}\right)$,
and, in the Kerr spacetime, asymptotically,
$U^{\alpha}\simeq\gamma\left(\delta_{0}^{\alpha}+v\delta_{r}^{\alpha}\right)$,
where $\gamma\equiv\sqrt{1-v^{2}}$. For this observer, the electromagnetic
tidal tensors are not symmetric:
$E_{\alpha\beta}=E_{(\alpha\beta)}+E_{[\alpha\beta]}$
and $B_{\alpha\beta}=B_{(\alpha\beta)}+B_{[\alpha\beta]}$, with:
\begin{eqnarray*}
E_{(\alpha\beta)}dx^{\alpha}dx^{\beta} & = &
\gamma\left(\frac{2q}{r^{3}}\left(vdt-dr\right)dr+\frac{q}{r}d\Omega_{2}-\frac{3\mu
v}{r^{3}}\sin^{2}\theta drd\phi\right)\\
E_{[\alpha\beta]} & = & -\frac{q\gamma v}{r^{3}}dr\wedge dt+\frac{3\mu\gamma
v}{2r^{2}}\left(\frac{\sin^{2}\theta}{r}dr\wedge d\phi-\sin2\theta
d\theta\wedge d\phi\right)\end{eqnarray*}
\begin{eqnarray*}
B_{(\alpha\beta)}dx^{\alpha}dx^{\beta} & = &
\frac{3\mu\gamma}{r^{2}}\left(\frac{2\cos\theta}{r^{2}}\left(vdt-dr\right)dr+\cos\theta
d\Omega_{2}+\frac{\sin\theta}{r}\left(vdt-2dr\right)d\theta\right)\\
B_{[\alpha\beta]} & = & \frac{q\gamma v\sin\theta}{r}d\theta\wedge
d\phi-\frac{3\mu\gamma v}{r^{3}}\left(\frac{\cos\theta}{r}dr\wedge
dt+\frac{\sin\theta}{2}d\theta\wedge dt\right)\end{eqnarray*}
while their gravitational counterparts are the symmetric tensors:
\begin{eqnarray*}
\mathbb{E}_{\alpha\beta}dx^{\alpha}dx^{\beta} & \simeq &
-\frac{2m\gamma^{2}}{r^{3}}\left(dr-vdt\right)^{2}+\frac{m}{r}d\Omega_{2}+\frac{6J\gamma^{2}v}{r^{3}}\sin^{2}\theta\left(vdt-dr\right)d\phi\\
&  &
+\mathcal{O}\left(r^{-4}\right)\left(dr+d\phi+dt\right)d\theta\end{eqnarray*}
\begin{eqnarray*}
\mathbb{H}_{\alpha\beta}dx^{\alpha}dx^{\beta} & \simeq &
\frac{3J}{r^{2}}\left(-\frac{2\gamma^{2}\cos\theta}{r^{2}}\left(dr-vdt\right)^{2}+\cos\theta
d\Omega_{2}+\frac{2\gamma^{2}\sin\theta}{r}\left(vdt-dr\right)d\theta\right)\\
&  &
+\left(\mathcal{O}\left(r^{-4}\right)dr+\mathcal{O}\left(r^{-3}\right)d\theta+\mathcal{O}\left(r^{-4}\right)dt\right)d\phi\end{eqnarray*}

Thus, while the static observer finds a similarity between gravitational
and electromagnetic tidal forces, for the observer moving relative
to the central body these forces are very different. Both observers will
however agree on the similarity between the invariants constructed
from gravitational and electromagnetic tidal tensors (see \ref{invk}).

\subsection{Ultra-stationary spacetimes}
\label{ultra}
 From the analysis of the previous section, one might conclude that a matching
between electromagnetic and gravitational tidal tensor components is
only possible in linearised theory. Indeed, the non-linear nature of the
curvature tensor would seem to preclude that an identification of metric
components with the components of the electromagnetic potential could yield
analogous tidal tensors, when expressed in terms of these components.
There is,
however, a class of spacetimes where the matching between electromagnetic and
gravitational tidal tensors is \textit{exact}. This class corresponds to
\textit{ultra-stationary spacetimes} defined as stationary spacetimes whose
metric has a constant $g_{00}$ component in the chart where it is explicitly
time independent. The most general metric for such spacetimes is
\bequ
ds^{2}=-\left(dt+A_{i}(x^k)dx^{i}\right)^{2}+\hat{g}_{ij}(x^k)dx^{i}dx^{j} \ .
\label{ultrastationary} \eequ
Consider now, in these spacetimes, the Klein-Gordon equation for a particle of
mass $m$: $\Box \Phi=m^2\Phi$; it reduces (with the ansatz
$\Phi=e^{-iEt}\Psi(x^j)$) to a time independent Schr\"odinger equation
$\hat{H}\Psi=\epsilon\Psi$, where
\bequ
\hat{H}=\frac{(\hat{\vec{P}}+E\vec{A})^2}{2m} \ , \ \ \ \
\epsilon=\frac{E^2-m^2}{2m} \ ; \label{Schrodinguer} \eequ
this is the non-relativistic problem of a ``charged'' particle with
charge $-E$
and mass $m$ under the influence of a magnetic field $\vec{B}=\nabla\times
\vec{A}$ in a curved space with metric $\hat{g}_{ij}$. This has been used to
map the Landau levels due to constant magnetic fields in three-spaces of
constant curvature to energy quantisation in spacetimes with Closed Timelike
Curves (like the G\"odel space) \cite{thesis,Drukker:2003mg}.

Computing now  the magnetic tidal tensor (\ref{Bab}) for the magnetic
field $\vec{B}=\nabla\times \vec{A}$ in a three space with metric
$\hat{g}_{ij}$:
\bequ
B_{ij}=\hat{D}_{j}B_{i}=\hat{\epsilon}_{lki}\hat{D}_j\hat{D}^lA^k \ . \label{ttemg}\eequ
where $\hat{D}$ denotes covariant differentiation carried out in the
three space
with metric $\hat{g}_{ij}$. This construction holds for the static observer of 4-velocity $U^{\mu}=\delta^{\mu}_0$; for the same observer, the magnetic
gravitational tidal tensor is \textit{exactly} given by
\bequ
\mathbb{H}_{ij}=\frac{1}{2}\hat{\epsilon}_{lki}\hat{D}_{j}\hat{D}^{l}A^{k} \ .
\label{tt} \eequ
Up to a factor of 2, $\mathbb{H}_{ij}$ and $B_{ij}$ are the same. This shows
that our interpretation of the magnetic part of the Riemann tensor
as a magnetic tidal tensor is indeed correct even outside the scope of
linearised theory. Turning the argument around, one can see this matching of
tidal tensors as a justification for the very close analogy between the
physics
in these apparently very different setups. In our approach this is to be
understood by the \textit{similarity of the tidal forces}.

It is somewhat surprising that, identifying some metric components with
electromagnetic potentials, the magnetic gravitational tidal tensor can have
exactly the same form as the magnetic tidal tensor, since the former would
normally be non-linear in the metric. Indeed, this is what happens to the
electric gravitational tidal tensor of (\ref{ultrastationary}): all its
components are non-linear in $A_i$, preventing an interpretation in
terms of an
electromagnetic field. Since there is no electric tidal tensor analogue, this
analogy is not perfect. But it is quite suggestive from the above mapping
between Klein-Gordon and Schr\"odinger equation that the magnetic part of the
problem  (together with the spacial curvature) captures all the physics.
Let us now exhibit three explicit examples of the analogy between the
gravitational magnetic tidal tensor of ultra-stationary spacetimes and the
magnetic tidal tensor of the analogous magnetic configuration.

\subsubsection{Som-Raychaudhuri, Van-Stockum and Heisenberg spacetimes}
\label{Van-Stockum}

The Som-Raychaudhuri metrics \cite{sr} are a family of solutions to the
Einstein-Maxwell equations with a source term representing charged incoherent
matter. They are characterised by two parameters, $a$ and $b$. The metric can
be written in the form (\ref{ultrastationary}) with
\bequ
A_{i}dx^{i}=-ar^{2}d\phi \ , \ \ \ \
\hat{g}_{ij}dx^idx^j=e^{-b^2r^{2}}dr^{2}+r^2d\phi^2+e^{-b^2r^{2}}dz^{2} \ .
\eequ
The special case with $a=b$ corresponds to the solution rediscovered in
1937 by
Van Stockum \cite{stockum} (first found by Lanczos \cite{lanczos}) describing
the metric in the interior of an infinitely long rotating cylinder of dust
(i.e., $p=0$) with energy density given by
$\mathbb{E}^{\alpha}_{\,\,\,\alpha}=4\pi\rho=2a^2\exp{a^2r^2}$.

The magnetic gravitational tidal tensor has non-vanishing components, for an
observer with 4-velocity $U^{\mu}=\delta^{\mu}_0$:
\bequ
\mathbb{H}_{rz}=\mathbb{H}_{zr}=-ab^{2}r \ . \label{stockum}\eequ
According to the Klein-Gordon equation, the analogous magnetic
configuration is
the magnetic field:
\begin{equation}
\vec{B}=\nabla\times\vec{A}=-2ae^{r^2b^2}\vec{e}_z
\end{equation}
living in a three space with metric $\hat{g}_{ij}$.
The tidal tensor of this field indeed has non-vanishing components $B_{rz}=B_{zr}=2\mathbb{H}_{rz}$.

If we put $b=0$, but keep $a\neq0$, the transverse space metric $\hat{g}_{ij}$
becomes flat. The four dimensional space becomes homogeneous, and the three
dimensional part $(t,r,\phi)$ is indeed a group manifold, the \textit{Nil} or
Heisenberg group manifold, which is the group associated to the Bianchi II Lie
algebra. The absence of tidal forces, in such case reflects the homogeneity of
the spacetime and of the analogous magnetic field. This group manifold is a
three dimensional version of the five dimensional maximally supersymmetric
G\"odel type universes found in $\mathcal{N}=2$ minimal ungauged five
dimensional supergravity in \cite{Gauntlett:2002nw}.

One other interesting point about this example is that the tidal tensors
$B_{\alpha\gamma}$ and $\mathbb{H}_{\alpha\gamma}$ are symmetric. According to
(\ref{MaxGrav}iv) this shows there are no mass currents. This is
counter-intuitive
for a rotating cylinder of fluid, unless the spacetime is described in
co-moving coordinates. Such coordinates correspond, precisely, to the
coordinate system above \cite{Mashhoon:1998fj}.

\subsubsection{The G\"odel Universe}
\label{Godel}

In 1949 Kurt G\"odel \cite{godel} found an exact solution of Einstein's field
equations which is usually portrayed as a homogeneous rotating universe. His
solution arose much discussion over the years due to the existence of Closed
Timelike Curves passing through any point which led G\"odel to recognise, and
explicitly discuss for the first time, that ``an observer travelling in this
universe could be able to travel to his own past''.

This description is conceptually hard. If the G\"odel universe is rotating and
is also homogeneous, then it must be rotating around \textit{any} point! What
does this mean? Of course one can argue that this solution should be discarded
as being unphysical, due to the CTC's. But even if such solution is unphysical
(which is not clear) building an intuition about it could be useful in
understanding General Relativity. We argue that the gravito-electromagnetic
analogy proposed herein helps us to construct such an intuition.
The metric can be written in the form (\ref{ultrastationary}) with
\bequ
A_{i}dx^{i}=e^{\sqrt{2}\omega x}dy \ , \ \ \ \ \hat{g}_{ij}dx^idx^j=dx^2+
\frac{1}{2}e^{2\sqrt{2}\omega x}dy^2+dz^{2} \ . \eequ
The magnetic gravitational tidal tensor vanishes, for an observer with
4-velocity $U^{\mu}=(1,\vec{0})$, which has led to conceptual difficulties in
literature (see section \ref{What physical conclusions}). The Klein-Gordon
equation maps this metric into a magnetic field:
\[
\vec{B}\equiv\nabla\times\vec{A}=2\omega\vec{e}_{z} \ , \]
on the three space with metric $\hat{g}_{ij}$. This field is uniform,
since the
magnetic tidal tensor $B_{\alpha\beta}$ vanishes. Thus, the physical
interpretation for the vanishing of $\mathbb{H}_{\alpha\beta}$ is that the
G\"odel universe, like the Heisenberg group manifold, has a \textit{uniform}
gravitomagnetic field.
The concept of homogeneous rotation can now be easily assimilated by an
analogy
with the more familiar picture of a gas of charged particles (e.g. the
conduction electrons in a metal) subject to a uniform magnetic field:
there are
Larmor orbits around any point.

\subsubsection{The metric of Lense and Thirring}
\label{Lense and Thirring}

In 1918, Lense and Thirring \cite{lense} solved the linearised Einstein
equations for a rotating thin spherical shell of matter, and found that
rotation can `drag inertial frames'. Since then, this \textit{Lense-Thirring
effect}  has become the archetype for magnetic effects in relativistic
gravity.
To linear order in $\omega$, the Lense-Thirring line element inside the
spherical shell is
\bequ
ds^2=-dt^2-2\omega r^2dtd\phi + dr^2+r^2d\phi^2+dz^2 \ .\label{ltmetric} \eequ
The metric can be written in the form
(\ref{ultrastationary}) with
\bequ
A_{i}dx^{i}=\omega r^{2}d\phi \ , \ \ \ \
\hat{g}_{ij}dx^idx^j=dr^{2}+(r^{2}+\omega^{2}r^{4})d\phi^2+dz^{2} \ . \eequ
It follows that the only non-trivial component of the magnetic gravitational
tidal tensor, for an observer with 4-velocity $U^{\mu}=(1,\vec{0})$, is
\bequ
\mathbb{H}_{zr}=-\frac{\omega^{3}r}{(1+r^{2}\omega^{2})^{3/2}}\ .
\label{lttt}\eequ
The tidal tensor is \textit{not} zero. Thus, the usual interpretation of this
spacetime as the gravitational analogue of a constant magnetic field (see e.g.
\cite{Tartaglia:2003wx}), must not be an exact assumption. To clear this
matter, we note, from (\ref{Schrodinguer}),
that the magnetic field configuration analogous to the metric is:
\[
\vec{A}=\frac{\omega}{1+\omega^{2}r^{2}}\vec{e}_{\phi} \ \ \
\Rightarrow \ \ \
\vec{B}=\frac{2\omega}{\sqrt{1+r^{2}\omega^{2}}}\vec{e}_{z}
\ , \]
which is not uniform (in the curved space with metric $\hat{g}_{ij}$),
since its
tidal tensor has the (only) non-vanishing component
$B_{zr}$=$2\mathbb{H}_{zr}$.

This is the correct interpretation for the magnetic analogue of the
Lense-Thirring metric if one takes it as an \textit{exact} metric.
However, for
small $r$ (and indeed the line element (\ref{ltmetric}) relates to the neighbourhood of the centre of the hollow sphere, cf. \cite{lense}), the uniform magnetic field picture is justified in the linear
approach, since $\mathbb{H}_{zr} \sim \mathcal{O}(\omega^3)$. Under these
conditions this spacetime is equivalent to the Heisenberg group
manifold, which
is exactly homogeneous and analogous to
a constant magnetic field in flat space.

\subsection{Force acting on a gyroscope}
\label{Stern-Gerlach}

The force acting on a dipole is a purely tidal effect; it is therefore
the most obvious physical application for a tidal tensor based analogy.
There is
no gravitational analogue to the electric dipole, since there are no
``negative
masses''; for the same reason, there can be no gravitational analogue to a
magnetic dipole ``alone''. But there is a clear gravitational analogue to the
electric pole - magnetic dipole particle, which is the ideal gyroscope
(i.e., a
pole-dipole spinning test particle, as defined in \cite{Papapetrou I}). In
gravity no force arises from the monopole term, since a spinless
particle moves
along a geodesic; hence, the force\footnote{The existence of such force causes
the gyroscope to deviate from geodesic motion; it well known, since the works
of Mathisson and Papapetrou \cite{Papapetrou I}, that a spinning particle
violates the weak equivalence principle.} exerted on a gyroscope should
indeed,
in the spirit of our approach, be the gravitational version of the
electromagnetic force exerted on a magnetic dipole.

In electromagnetism, the force acting on a magnetic dipole when placed
in a magnetic field $\vec{B}$ is usually given in literature (see, for
instance,
\cite{Gravitation and Inertia}, p. 318)  by:
\begin{equation}
\vec{F}_{EM}=\nabla(\vec{\mu.}\vec{B})\
\label{ForceEMnCov} \eequ
where $\vec{\mu}=\frac{q}{2m}\vec{S}$ denotes the magnetic dipole
moment, and $\vec{S}$ the classical angular momentum. A covariant form of this
equation is:
\begin{equation}
F_{EM}^{\beta}=\frac{q}{2m}B^{\alpha\beta}S_{\alpha}\ ,
\label{ForceEM} \eequ
where $B_{\alpha\beta}=F_{\alpha\gamma;\beta}U^{\gamma}$ is the magnetic
tidal tensor {}``seen'' by the dipole of 4-velocity $U^{\gamma}$,
and $S^{\alpha}$ is the ``intrinsic angular momentum''%
(e.g. \cite{Gravitation}, p. 158), defined as being the 4-vector with
components
$(0,\vec{S})$ in the dipole rest
frame. Equation (\ref{ForceEM}) is valid for a dipole moving with arbitrary
velocity; hence, from the definition of the magnetic tidal tensor (\ref{Bab}),
and a simple covariantization of the non-relativistic expression
(\ref{ForceEMnCov}), we have just obtained the important equation of the
electromagnetic force exerted on a moving magnetic dipole, avoiding an
otherwise more demanding computation (eg. \cite{Stephani}, pp. 80-84).

In the light of our approach, the analog force in gravity should then
be:
\begin{equation}
F_{G}^{\beta}=-\mathbb{H}^{\alpha\beta}S_{\alpha}\ ,
\label{ForceGR} \eequ
where
$\mathbb{H}_{\alpha\beta}=R_{\alpha\gamma\beta\delta}U^{\gamma}U^{\delta}$
is
the gravitational magnetic tidal tensor {}``seen'' by the gyroscope
of 4-velocity $U^{\gamma}$ and intrinsic angular momentum $S^{\alpha}$. The
minus sign reflects the fact that two masses ``of the same sign''
\textit{attract} each other, by contrast with electromagnetism, where two
charges of same sign \textit{repel} each other; this implies that mass
currents
in the same direction \textit{repel} each other, while charge currents in the
same direction \textit{attract} each other (see \cite{Schutz} p. 246 and
\cite{Feynman} 13-6 for an excellent explanation of these analogue effects).
The factor $q/(2m)$ in (\ref{ForceEM}) also drops out since the gravitational
analog of $\vec{\mu}$ is $\vec{S}$, as can be seen by comparing
(\ref{MaxGrav}iv) to (\ref{MaxCov}iv) (see also discussion in section
\ref{rotatingp}).

Equation (\ref{ForceGR}) turns out to be the exact result
derived by Papapetrou \cite{Papapetrou I}-\cite{Wald Stern
Gerlach}\cite{Harris1991,Gravitation and Inertia,Stephani,Gravitation}:
\begin{equation}
\frac{DP^{\alpha}}{D\tau}=-\frac{1}{2}R_{\,\,\,\beta\mu\nu}^{\alpha}U^{\beta}S^{\mu\nu}\label{Papapetrou}\end{equation}with
Pirani \cite{Pirani 1956} supplementary condition $S^{\mu\nu}U_{\nu}=0$, where
$S^{\mu\nu}$ denotes the spin tensor (e.g. \cite{Gravitation}, p. 158). This
result may therefore be seen as a definite
confirmation that our interpretation of $\mathbb{H}_{\alpha\beta}$
as a magnetic tidal tensor is indeed correct.

This gravito-electromagnetic analogy based derivation of the Papapetrou
equation
has two major strengths. The first, its obvious simplicity:
(\ref{ForceGR}) was
obtained in a two step derivation, avoiding the lengthy original computation
\cite{Papapetrou I}. The second is that, when written in a form explicitly
analogue to its electromagnetic counterpart (\ref{ForceEM}), it makes
possible,
by comparison with the more familiar electromagnetic ones, to visualise
effects
which are
not transparent at all in the form (\ref{Papapetrou}) presented in literature.
At the same time, it also reveals in a clear fashion significant differences
between the electromagnetic
and gravitational forces, which arise from the different symmetries
of the tidal tensors. In particular, due to these symmetries, as follows from
equations (\ref{MaxCov}iv) and (\ref{MaxGrav}iv), the two forces can only be
similar when the fields are static (besides weak) in the dipole rest frame.
This will be discussed in detail in \cite{gravitational stern-gerlach}.
Herein, we will just focus on the temporal components of these forces. Let us
start by the electromagnetic force (\ref{ForceEM}); its temporal component is
not (generically) zero in the dipole rest frame, as one might naively
expect\footnote{Consider, as an example, the tensorial form of the Lorentz
Force: $DP^{\alpha}/D\tau=qF^{\alpha\beta}U_{\beta}$. It also exhibits a time
component, given by $DP^{0}/D\tau=\vec{v}.\vec{E}/\sqrt{1-v^{2}}$, which
represents the work done, per unit time, by the electromagnetic field on the
particle of charge $q$. It is zero in the particle rest frame, since the
particle (obviously) does not move relative to itself. But that is not
the case
of the time component of force (\ref{ForceEM}), which depends on
quantities that
are all measurable in the dipole rest frame.}. In the dipole rest frame we
have:\[F^{0}_{EM}=B^{i0}S_{i}\frac{q}{2m}=B^{i0}\mu_{i}=-\frac{\partial
B^{i}}{\partial t}\mu_{i}\]The magnetic dipole may be seen as a small current
loop; denoting the area of the loop by $a$, and its current by $I$, the
magnetic dipole moment is then given by $\vec{\mu}=IA\vec{n}$, where $\vec{n}$
is the vector normal to the plane of the loop. Thus:
\[F_{EM}^{0}=-\frac{\partial\vec{B}}{\partial
t}.\vec{n}AI=-\frac{\partial\Phi}{\partial
t}.I=I\oint_{loop}\vec{E}.\vec{ds}=P\]where $\Phi\equiv$ \{magnetic flux
through the loop\}, $\vec{E}\equiv$ \{induced electric field\} and $P\equiv$
\{net work done per unit time by the the induced electric forces\}. This way,
in the dipole rest frame, the time component of (\ref{ForceEM}) gets a simple
physical interpretation: it is the power transferred to the dipole by
electromagnetic Maxwell-Faraday induction, due to a time varying magnetic
field.

We turn now to gravity. Unlike its electromagnetic counterpart,
$\mathbb{H}_{\alpha\beta}$ is a spacial tensor; this means that if
$\mathbb{H}_{\alpha\beta}$ is the gravitational magnetic tidal tensor as
measured by a given observer of 4-velocity $U^{\gamma}$, then
$\mathbb{H}_{\alpha\beta}U^{\alpha}=\mathbb{H}_{\alpha\beta}U^{\beta}=0$. Or,
equivalently, that, in the coordinates of the observer's rest frame,
$\mathbb{H}_{\alpha\beta}$ has only space components. Therefore, in the
gyroscope rest frame:\[F^{0}_{G}=-\mathbb{H}^{i0}S_{i}=0\]and one may thus
regard the spatial character of the gravitational tidal tensors as
evidence for
the \textit{nonexistence of electromagnetic-like induction effects in
gravity},
which is in accordance with discussion in section \ref{Gravity vs
Electromagnetism}.

We close this section with a remark on how this application emphasises the
universality of the gravito-electromagnetic analogy based on tidal tensors:
gravitational and electromagnetic tidal tensors will not be similar except
under very special conditions; generically, $B_{\alpha\beta}$ and
$\mathbb{H}_{\alpha\beta}$ will not even exhibit the same symmetries -
but they
will always play analogous roles in dynamics, as shown by equations
(\ref{ForceEM}) and (\ref{ForceGR}).

\section{Linearised theory approach}
\label{linearth} We started our analogy by looking under which circumstances
tidal effects could have a similar tensorial description in gravity
and electromagnetism. We have seen (section \ref{minper}) that the similarity
between tidal
tensors we found on certain special conditions could be regarded as
an analogy between some components of the metric tensor and the
electromagnetic
gauge potentials.

Such an analogy is also suggested by the formal similarity between
the linearised Einstein equations in the harmonic gauge, and the Maxwell
equations in the Lorentz Gauge. In the linearised theory approach
to gravitoelectromagnetism (GEM), this analogy is worked out to define the
gravitational analogues
to the electromagnetic fields $\vec{E}$ and $\vec{B}$, with which
one aims to describe gravity. Of course, since gravity is pure geometry,
such forces or fields have no place in it; in this sense, we
may see this approach as an analogy between physical quantities which
are present in one theory, but do not exist in the other. This marks
an important conceptual difference from the approach based on tidal
tensors, where we compare quantities common to both theories.

Nevertheless, we know that in the Newtonian limit (i.e., when the field
is weak, and the relative motion of the sources is negligible), the
linearised Einstein equations reduce to the Poisson equation, and
gravity is well described by the Newtonian gravitational field. We
also know that when the motion of the sources is taken into account,
magnetic-type phenomena occur in gravity; and given the aforementioned
similarity between the linearised Einstein equations in the harmonic
gauge and the Maxwell equations in the Lorentz gauge, it is legitimate
to suppose that gravity can be described, in some appropriate limit,
by fields analogous to the electromagnetic ones.

There is no general agreement about the limit of validity of this
analogy (see, for instance, \cite{Harris1991} and \cite{Harmonic
Gauge}), apart
from the fact that it is defined in the weak field and slow motion
approximation. While some authors limit the analogy to stationary
configurations
\cite{Harris1991}-\cite{General Relativity},
for others it can be extended to time dependent setups
\cite{Mashhoon:2003ax}-{\cite{Harmonic Gauge}.

We argue that our analysis of the tidal tensors sheds light into this
debate: gravity can be described, in an approximate way, by fields
analogue to those from electromagnetism only when the gravitational
tidal tensors are similar to their electromagnetic counterparts (i.e.,
when the tidal forces, which are the only forces present in gravity,
are correctly determined by the variations of those fields). In particular,
this requires the fields to be static in the observer's rest frame, as
is clear
from the comparison of equations (\ref{MaxCov}) with (\ref{MaxGrav}) (see also
section \ref{minper}).
Otherwise, gravitational tidal tensors do not even exhibit the same symmetries
as the covariant derivative of an electromagnetic type field, signalling that
such
fields are no longer appropriate to describe gravity.

We shall now briefly review and discuss these two versions of linear
approach to
GEM.

\subsection{Time independent approach}
\label{standard approach}
One starts by considering small perturbations $h_{\mu\nu}$ around
Minkowski spacetime, such that the metric is given
by:
\begin{equation}
g_{\mu\nu}=\eta_{\mu\nu}+h_{\mu\nu},\quad |h_{\mu\nu}|\ll1 \label{linearmetric}
\end{equation}
 It is more convenient
to work with the quantity $\bar{h}_{\mu\nu}\equiv
h_{\mu\nu}-\eta_{\mu\nu}h_{\,\,\alpha}^{\alpha}/2$;
imposing a gauge condition called \textit{de Donder or harmonic gauge},
given by $\bar{h}_{\alpha\beta}^{\,\,\,\,\,\,,\beta}=0$, the linearised
Einstein equations take the form\footnote{In this section, following the
standard procedure in the literature, we reinsert the speed of light
$c$, while
still making $G=1$.}:
\begin{equation}
\square\bar{h}^{\alpha\beta}=-\frac{16\pi}{c^{4}}T^{\alpha\beta}\label{Einstein}\end{equation}

We will consider a metric with general perturbations of the form (\ref{mper}), but with $g_{ij}=\delta_{ij}$, i.e., using \textit{nearly Lorentz coordinates}, so that conditions (\ref{linearmetric}) are satisfied. For such metric the harmonic gauge condition yields:
\begin{equation}
\frac{1}{c^{3}}\frac{\partial}{\partial
t}\left(\phi+3C\right)+\frac{2}{c^{2}}\nabla.\vec{A}=0,\quad\frac{2}{c^{3}}\frac{\partial
A_{i}}{\partial
t}+\frac{1}{c^{2}}\left[\left(\phi-C\right)_{,i}+2\xi_{ij}^{\,\,\,,j}\right]=0\label{harmonic
1}\end{equation}
In the special case $C(t,x^{k})=\phi(t,x^{i}),\:  \xi_{ij}(t,x^{k})=0$
one gets $\bar{h}_{ij}=0$ and (\ref{Einstein}) reduces to the set
of four equations for $\bar{h}^{0\beta}$ which, apart
from the different tensorial structure, exhibit a clear analogy with
the Maxwell equations in the Lorentz gauge: \[
\square A^{\beta}=-\frac{4\pi}{c} j^{\beta}\]
Furthermore, the harmonic gauge condition (\ref{harmonic 1}) reduces
to:
\begin{equation}
\frac{1}{c^{3}}\frac{\partial\phi}{\partial
t}+\frac{\nabla.\vec{A}}{2c^{2}}=0,\quad\frac{1}{c^{3}}\frac{\partial\vec{A}}{\partial
t}=0 \label{harmonic Harris}
\end{equation}
The second equation imposes a static potential $\vec{A}$,
while the first one is similar (up to a factor of 2 in the second
term) to the Lorentz gauge condition in electromagnetism; thus, defining
the gravitoelectric and gravitomagnetic fields in direct analogy with
electromagnetism:
\begin{equation}
\vec{E}_{G}\equiv-\nabla\phi,\qquad\vec{B}_{G}\equiv\nabla\times\vec{A}\label{Def
campos}\end{equation}
these two definitions, together with equations (\ref{Einstein}) and
(\ref{harmonic Harris}) lead to the following set of equations
\cite{Harris1991}:\begin{eqnarray}
\mbox{i)}\,\nabla.\vec{E}_{G}=4\pi\rho_{m} &  &
\mbox{ii)}\,\nabla\times\vec{E}_{G}=0\nonumber \\
\mbox{iii)}\,\nabla.\vec{B}_{G}=0\quad\,\, &  &
\mbox{iv)}\,\frac{1}{2}\nabla\times\vec{B}_{G}=\frac{4\pi}{c}\vec{J}+\frac{1}{c}\frac{\partial\vec{E}_{G}}{\partial
t}\label{eqs Harris}\end{eqnarray}
The last equation would generally imply $\nabla\times\vec{B}_{G}\ne0$
in empty space, what seems to contradict our exact equation (\ref{MaxGrav}iv),
which states that the magnetic gravitational tidal tensor is always
symmetric in empty space; to clear this matter we will check the geodesics.
To linear order in the potentials and velocity of the test particle,
the geodesic equation yields:
\begin{equation}
\frac{d^{2}\vec{x}}{dt^{2}}=-\vec{E}_{G}-\frac{2}{c}\vec{v}\times\vec{B}_{G}-\frac{2}{c^{2}}\frac{\partial\phi}{\partial
t}\vec{v}\label{geodesicas Harris}\end{equation}
which reduces to the form (\ref{geolorentz}) analogue to the electromagnetic
Lorentz force
\emph{only} if $\partial\phi/\partial t=0$; thus, this analogy between
electromagnetism and linearised gravity holds only if the potentials
are \textit{static}, as is asserted in \cite{Gravitation and Spacetime}
p. 163,
and in accordance with our discussion in section \ref{minper}. In this case,
the last term of (\ref{eqs Harris}iv) vanishes, and equations (\ref{eqs
Harris}) become but an approximation, in the
limit of static weak fields and stressless sources%
\footnote{The source of the gravito-electric field is here $\rho_{m}$ instead
of $2\rho_{m}+T_{\,\,\alpha}^{\alpha}$ which appears in our equation
(\ref{MaxGrav}i). This physical content is lost here since, by choosing
$\phi=C$ and $\xi_{ij}=0$ above, we imposed $\bar{h}_{ij}=0$;
to be valid everywhere, this condition indeed demands, by virtue of
equations (\ref{Einstein}), $T_{ij}=0$, so that
$T_{\,\,\alpha}^{\alpha}=-\rho_{m}$. This is consistent with the slow-source
assumption, where the pressure and all material stresses are taken to be
negligible compared to the energy density term.}, to our exact equations
(\ref{MaxGrav}).

The non-covariance of this analogy (already manifest in its formalism) is
obvious: if one considers a Lorentz boosted observer, the analogy will no
longer hold since, for that observer, the potentials will be time dependent.

Nevertheless, despite its limitations, this approach still proves extremely
useful whenever it applies. The gravitomagnetic fields (\ref{Def campos}) are the same as (\ref{E B nossa analogia}) for the special case $g_{ij}=\delta_{ij}$; thus, cf. discussion in section \ref{minper}, they help us visualise frame
dragging, and lead directly to the Lense and Thirring precession of (for
instance) a test gyroscope (in the units system used in this section, the
torque applied on the gyroscope is $\tau=-\vec{S}\times\vec{B}_{G}/c$, which
yields the precession frequency $\vec{\omega}=\vec{B}_{G}/c$).
The gravito-magnetic field $\vec{B}_G$ has also been used \cite{Harris1991,
Gravitation and Inertia, Wald Stern Gerlach} to obtain a first order estimate
to the Papapetrou force applied on the gyroscope (\ref{ForceGR}); in direct
analogy with the electromagnetic expression (\ref{ForceEMnCov}), that first
order estimate is $\vec{F}_{G}=-\nabla(\vec{S.}\vec{B_{G}})/c$. However, as is
asserted in \cite{Wald Stern Gerlach}, and made clear by the discussion in
section \ref{Stern-Gerlach}, this expression is valid only when the gyroscope
is at \textit{rest} in a \textit{stationary} spacetime; otherwise, the
gravitational and electromagnetic forces will differ significantly, due to the
different symmetries of the tidal tensors ``seen'' by the gyroscope.

\subsection{``Maxwellian'' gravity}
\label{Maxwellian gravity}
On some of the literature covering
the linear perturbation approach to GEM, a set of ``gravitational
Maxwell equations'' is derived
\cite{Mashhoon:2003ax}-\cite{Mashhoon:1999nr}\cite{Forward
Antigravity}-\cite{Harmonic Gauge}, which \textit{include} time dependent
terms; these equations differ from the set (\ref{eqs Harris}) by the
following:
the last term of (\ref{eqs Harris}iv) is now kept, and (\ref{eqs Harris}ii) is
replaced by:
\begin{equation}
\nabla\times\vec{E}_{G}=-\frac{1}{2c}\frac{\partial\vec{B}_{G}}{\partial
t}\label{Maxwelliana}\end{equation}
Based on this kind of equations, some exotic gravitational phenomena
have been predicted, such as the existence of a gravitational analogue
of Faraday's law of induction \cite{Tartaglia:2003wx,Forward
Antigravity,Thorne
Braginsky},
deduced from (\ref{Maxwelliana}).
Experiments to detect it have been proposed \cite{Thorne Braginsky}
and, recently, such an experiment has actually been performed \cite{ESA}.

These equations have been under debate\footnote{To follow this debate, see
\cite{Thorne Braginsky}, \cite{Harris1991}, \cite{Gravitation and
Spacetime} p.
163, and \cite{Harmonic Gauge}.} in literature, and there is
no consensus around them;
they are also in contradiction with the results from our approach,
since in dynamical situations they would imply tidal tensors having
symmetries that, by virtue of our exact, covariant equations (\ref{MaxGrav}ii)
and (\ref{MaxGrav}iv), gravitational tidal tensors can never have.

Note that the physical content of the equations depends crucially
on these symmetries. For instance, in the framework of the approach
based on tidal tensors, it is clear that such phenomena as an analogue
of Faraday's induction cannot take place in gravity: our exact equation
(\ref{MaxGrav}ii) shows that the electric tidal tensor is always
symmetric, i.e., the {}``gravito-electric field'' is (always) irrotational.

In order to clarify these contradictions, let us
review the derivation of these Maxwell-like equations. One starts by
making $C(t,x^{k})=\phi(t,x^{i})$ in (\ref{mper}), then
one neglects all terms of order $c^{-4}$ or lower; that amounts to
neglect (see eg. \cite{Mashhoon:1999nr}) the tensor perturbations
term in (\ref{mper}) since in the slow motion approximation
$\xi_{ij}(t,x^{k})\sim c^{-2}$. The next step is to define the gravitational
electric $\vec{E}_{G}$ and magnetic $\vec{B}_{G}$ fields; the latter
is defined as in (\ref{Def campos}), while the former is here:\begin{equation}
\vec{E}_{G}\equiv-\nabla\phi-\frac{1}{2c}\frac{\partial\vec{A}}{\partial
t}\label{E Mashhoon}\end{equation}
This definition leads directly to equation (\ref{Maxwelliana}). The other
equations are obtained using these definitions together with equations
(\ref{Einstein}) and (\ref{harmonic 1}).

In a consistent approximation, however, the last term of (\ref{E Mashhoon})
should have been dropped as in (\ref{Def campos}), since by virtue
of the second equation in (\ref{harmonic 1}), neglecting
tensor perturbations amounts to neglect the time dependence of $\vec{A}$ (cf.
\cite{General Relativity}, p. 78). This condition is however ignored, again
with the argument (see eg. \cite{Mashhoon:1999nr}) that it involves terms
of order\footnote{Note, from eqs. (\ref{Einstein}), that in the slow motion
approximation, $|\vec{A}|\sim c^{-1}$.} $c^{-4}$; and this procedure
is indeed the origin of the controversy surrounding this particular
approach (see eg. \cite{Harris1991} and \cite{Harmonic Gauge}). In what
follows
we will show that indeed this analogy breaks down precisely when time
dependent
terms are present.

An analysis of the tidal effects reveals quite clearly that definition
(\ref{E Mashhoon}) cannot be appropriate for a gravitational field.
According to (\ref{E Mashhoon}) an observer at rest would be subject
to tidal forces described by a tensor of the form:
\begin{equation}
(E_{G})_{i,j}=-\phi_{,ij}-\frac{1}{2c}\frac{\partial A_{i,j}}{\partial
t}\label{tensor Mashhoon}\end{equation}
which is not symmetric when $\partial\vec{A}/\partial t\ne0$. But
we know, from the geodesic deviation equation, that the tidal tensor,
for this observer, is actually given, to linear order, by:
\begin{equation}
\mathbb{E}_{ij}=R_{i\beta
j\gamma}\delta_{\,\,0}^{\beta}\delta_{\,\,0}^{\gamma}c^{2}\approx-\phi_{,ij}-\frac{2}{c}\frac{\partial
A_{(i,j)}}{\partial t}-\frac{1}{c^{2}}\frac{\partial^{2}\phi}{\partial
t^{2}}\delta_{ij}\label{tensor mares linearizado}\end{equation}
which is symmetric as expected; note that the second term in (\ref{tensor
Mashhoon}),
as well as both the second and third terms in (\ref{tensor mares linearizado})
are of same order $c^{-2}$, and cannot be consistently neglected in
a approximation where terms down to $c^{-3}$ are kept. Therefore,
within this approximation, the only way to make expressions (\ref{tensor
Mashhoon})
and (\ref{tensor mares linearizado}) match is to consider static
potentials (where both reduce to the Newtonian tidal tensor).

An inspection of the geodesics leads to a similar conclusion. To linear
order in the potentials and velocity of the test particle, the geodesic
equation yields:
\[
\frac{d^{2}\vec{x}}{dt^{2}}=-\vec{E}_{G}+\frac{1}{c}\left(\frac{3}{2}\frac{\partial\vec{A}}{\partial
t}-2\vec{v}\times\vec{B}_{G}\right)-\frac{2}{c^{2}}\frac{\partial\phi}{\partial
t}\vec{v}\]
in this equation all terms apart from $\vec{E}_{G}$ are of same
order $c^{-2}$; therefore, in this approximation, the only way to
obtain a Lorentz like expression (\ref{geolorentz}), is to consider static
potentials \cite{Gravitation and Spacetime}-\cite{Ruggiero:2002hz}\cite{Wald
Stern Gerlach}.

As a third evidence that gravity cannot be described by electromagnetic-like
fields when the spacetime is not stationary, we point the fact that, on those
conditions, the gravito-magnetic field $\vec{B}_{G}$ does not yield the
correct
force applied on a gyroscope. According to the literature where this
``Maxwellian'' analogy is pursued, that force would be given by (see eg.
\cite{Mashhoon:2003ax,Ruggiero:2002hz,Mashhoon:1999nr}):
\begin{equation}
\vec{F}=-\frac{1}{c}\nabla(\vec{S.}\vec{B_{G}})\quad \Leftrightarrow \quad
F^{j}=-\frac{1}{c}(B_{G})^{i,j}S_{i}\label{ForceTL}
\eequ
But, as already discussed in previous section, the validity of this equation
requires the field to be static \cite{Harris1991, Gravitation and
Inertia, Wald
Stern Gerlach}. And it is easy to show it leads to an incorrect result when
applied to time dependent fields. The correct force is given by the Papapetrou
equation (\ref{ForceGR}), which, when applied to a gyroscope at rest in the
above considered spacetime reads, to linear order:
\begin{equation}
F_{G}^{j}=-\frac{1}{c}\mathbb{H}^{i j}S_{i}\approx
-\frac{1}{c}\left[(B_{G})^{i,j}-\frac{1}{c}\frac{\partial}{\partial
t}\left(\epsilon^{ijk}\phi_{,k}\right)\right]S_{i}
\label{ForceTLcorrecta}
\eequ
This equation only matches (\ref{ForceTL}) if the potentials are static
(again,
both terms in (\ref{ForceTLcorrecta}) are of same order in powers of $c^{-1}$;
hence, none can be neglected). This way we see that, even within the
scope of a
linear perturbation approach, one must use the analogy based on tidal tensors
instead of gravito-electromagnetic fields in order to obtain the
correct result
when the spacetime is not stationary.

Thus, to conclude and summarise: we argue that the physical analogy
$\{\vec{E},\vec{B}\}\leftrightarrow \{\vec{E}_{G},\vec{B}_{G}\}$ must be
restricted to time
independent setups, as is done in most textbooks covering the subject
(e.g. \cite{Gravitation and Spacetime,Gravitation and Inertia,General
Relativity}), and in accordance with the results from
our approach based on tidal tensors.

\section{The analogy between gravitational \textit{tidal tensors} and
electromagnetic \textit{fields}}
\label{secanalogy2}
There is a different analogy \cite{matte}-\cite{Ferrando:03} between Maxwell's
equations of electromagnetism and
the so called ``higher order'' gravitational field equations which has led to
the definition of an electric and a magnetic part of the curvature tensor,
originally introduced in
\cite{matte} (see also \cite{bel}). These electric and magnetic parts of the
curvature tensor form, moreover, invariants in a similar fashion to the
relativistic invariants formed by the electric and magnetic fields
\cite{matte,Bonnor:1995zf,cherubini:02,Maartens:1997fg}. In this
section we will review this analogy and dissect its physical content in the
light of our approach based on tidal tensors.

\subsection{General arguments}
\label{secanalogy General Arguments}
Maxwell's field equations in vacuum are
\bequ
F^{\mu}_{\  \nu;\mu}=0 \ , \ \ \ \ \ F_{[\mu\nu;\alpha]}=0 \ ,
\label{fe} \eequ
where the second set of equations are, in fact, Bianchi identities. Given a
congruence of observers with $4$-velocity field $U^{\alpha}$, we can split the
Maxwell tensor into the two spatial vector fields:
\bequ
E^{\alpha}=F^{\alpha}_{\  \beta}U^{\beta} \ , \ \ \ \ \ B^{\alpha}=\star
F^{\alpha}_{\ \ \beta}U^{\beta} \ , \label{split} \eequ
which are, of course, the electric and magnetic fields as measured by the
observer of 4-velocity $U^{\alpha}$, but in this context dubbed
\textit{electric and magnetic
parts of the Maxwell tensor}.  They completely characterise the latter, as can
be
seen by writing \bequ
F_{\alpha\beta}=2U_{[\alpha}E_{\beta]}+\epsilon_{\alpha\beta\mu
\nu}B^{\mu}U^{\nu} \ . \label{decompF}\eequ
Thus, all 6 independent components of $F_{\mu \nu}$ are encoded in the 3+3
independent components of $E^{\alpha}$ and $B^{\alpha}$ and there is an
equivalence between the vanishing of $F_{\alpha \beta}$ and the simultaneous
vanishing of $E_{\alpha}$ and $B_{\alpha}$. In spite of their dependence on
$U^{\alpha}$, one can use $E_{\alpha}$ and $B_{\alpha}$ to define two
tensorial
quantities which are $U^{\alpha}$ independent, namely
\bequ
E^{\alpha}E_{\alpha}-B^{\alpha}B_{\alpha}=-\frac{F_{\alpha \beta}F^{\alpha
\beta}}{2} \ , \ \ \ \ \ E^{\alpha}B_{\alpha}=-\frac{F_{\alpha \beta}\star
F^{\alpha \beta}}{4} \ ; \label{inv} \eequ
these are the two independent relativistic invariants in four spacetime
dimensions.

Let us now turn to gravity. The curvature tensor obeys the second Bianchi
identity $R_{\sigma\tau[\mu\nu;\alpha]}=0$. In vacuum, these are equivalent to
\bequ
R^{\mu}_{\  \nu\sigma\tau;\mu}=0 \ , \label{hofe}\eequ
by virtue of the field equations $R_{\mu \nu}=0$. Let us change the
perspective
by observing the following result (originally due to Lichnerowicz, see
\cite{bel}): Let $\Sigma$ be a spacelike hypersurface and $V$ a neighbourhood
of $\Sigma$; then, if $R^{\mu}_{\  \nu\sigma\tau;\mu}=0$ in $V$, it follows
from $R_{\sigma\tau[\mu\nu;\alpha]}=0$ that $R_{\mu \nu}=0$ in $V$ iff $R_{\mu
\nu}=0$ in $\Sigma$. Thus, $R_{\mu \nu}=0$ becomes an initial condition in
$\Sigma$ which is propagated to $V$ by virtue of the \textit{higher
order field
equations} (\ref{hofe}) and the Bianchi identities. Taking this
perspective the
gravitational analogue of (\ref{fe}) are
\bequ
R^{\mu}_{\  \nu\sigma\tau;\mu}=0 \ , \ \ \ \ \
R_{\sigma\tau[\mu\nu;\alpha]}=0 \
. \label{fe2} \eequ
 From decomposition (\ref{Riemann-Weyl}) we see that in vacuum:
$R_{\alpha\beta\gamma\delta}=C_{\alpha\beta\gamma\delta}$. Again, given a
congruence of observers with $4$-velocity field $U^{\alpha}$, we
can split the curvature tensor into its electric and magnetic parts given in
(\ref{Weyl E H}). These two spatial tensors, both of
which are symmetric and traceless, completely characterise the Weyl tensor,
as can be seen by writing
\bequ
C_{\alpha\beta}^{\ \ \ \gamma
\delta}=4\left\{2U_{[\alpha}U^{[\gamma}+g_{[\alpha}^{\
[\gamma}\right\}\mathcal{E}_{\beta]}^{\
\delta]}+2\left\{\epsilon_{\alpha \beta
\mu \nu}U^{[\gamma}\mathcal{H}^{\delta]\mu}U^{\nu}+\epsilon^{\gamma \delta \mu
\nu}U_{[\alpha}\mathcal{H}_{\beta]\mu}U_{\nu}\right\} \ . \label{decomp3}\eequ
Thus, all 10  independent components (in vacuum) of $R_{\mu \nu \alpha \beta}$
are encoded in the 5+5 independent components of
$\mathcal{E}_{\alpha\beta}$ and
$\mathcal{H}_{\alpha\beta}$. Again, in spite of their dependence on
$U^{\alpha}$,
one can use $\mathcal{E}_{\alpha\beta}$ and $\mathcal{H}_{\alpha\beta}$ to
define
two tensorial quantities which are $U^{\alpha}$ independent, namely

\bequ
\mathcal{E}^{\alpha\beta}\mathcal{E}_{\alpha\beta}-\mathcal{H}^{\alpha\beta}\mathcal{H}_{\alpha\beta}=\frac{C_{\alpha
\beta\mu \nu}C^{\alpha \beta\mu \nu}}{8} \ , \ \ \ \ \
\mathcal{E}^{\alpha\beta}\mathcal{H}_{\alpha\beta}=\frac{C_{\alpha \beta\mu
\nu}(\star C)^{\alpha \beta\mu \nu}}{16} \ . \label{inv3}\eequ
The construction (\ref{fe2})-(\ref{inv3}) is, clearly, formally analogous to
(\ref{fe})-(\ref{inv}).
Let us now consider electromagnetic/gravitational sources. Maxwell's field
equations (\ref{fe}) get replaced by
\bequ
F^{\mu}_{\  \nu;\mu}=J_{\nu} \ , \ \ \ \ \ F_{[\mu\nu;\alpha]}=0 \ ,
\label{fe3}
\eequ
where $J_{\nu}$ is the $4$-current. Since including sources does not
change the
number of independent components of the Maxwell tensor, we can still split it
as in (\ref{split}), so that the remaining construction
(\ref{split})-(\ref{inv}) is unchanged. The same does not apply,
however, to the
Riemann tensor: in the presence of sources it cannot be completely
characterised
by its electric and magnetic parts. The reason can be seen as follows.
Including
sources endows the Riemann tensor with non-vanishing trace and thus
with its maximal number of independent components: 20 in four
dimensions. On the
other hand, from equations
(\ref{ERiemann-EWeyl})-(\ref{HRiemann-HWeyl}) we see
that the sources endow $\mathbb{E}_{\alpha\beta}$ with a trace and
$\mathbb{H}_{\alpha\beta}$ with an anti-symmetric
part; then, these tensors combined possess at most 6+8 components, which is
insufficient to encode all the information in
the Riemann tensor.

Decomposition (\ref{decomp3}), in its turn, remains valid in the presence of
sources, since the Weyl tensor is by definition traceless and has generically
only 10 independent components in four dimensions, which are completely
encoded
into the 5+5 components of $\mathcal{E}_{\alpha\beta}$ and
$\mathcal{H}_{\alpha\beta}$. For this reason, generically, in this
construction
one uses $C_{\alpha\beta\gamma\delta}$ instead of
$R_{\alpha\beta\gamma\delta}$.
Hence, we replace (\ref{fe2}) by\footnote{Note that
the two equations in (\ref{fe4}) are equivalent, using the usual Einstein
equations, unlike the two equations in (\ref{fe3}). But the perspective
in this
analogy is that these are the fundamental equations, and the usual Einstein
equations are initial conditions propagated to the whole spacetime by
virtue of
these equations.}
\bequ
C^{\mu}_{\
\nu\sigma\tau;\mu}=T_{\nu[\tau;\sigma]}-\frac{1}{3}g_{\nu[\tau}T_{;\sigma]} \
,
\ \ \ \ \ R_{\sigma\tau[\mu\nu;\alpha]}=0 \ , \label{fe4} \eequ
where $T_{\mu \nu}$ is the energy momentum tensor, with trace $T$, and we have
used the Einstein equations in the form $G_{\mu \nu}=T_{\mu \nu}$. Equations
(\ref{fe4}), together with (\ref{decomp3})-(\ref{inv3}) form, in the spirit of
this approach, the general gravitational analogue to construction
(\ref{split})-(\ref{inv}).

Generically we can say that the electric and magnetic parts of the Weyl tensor
describe the \textit{free} gravitational field, since they correspond to the
parts of the curvature that do not couple directly to sources, but rather just
via the integrability conditions (i.e. the Bianchi identities). In 3 spacetime
dimensions, the curvature is completely determined by the sources; thus there
is no free gravitational field. That is why the Weyl tensor is
identically zero
in 3 spacetime dimensions.

\subsection{The analogy in the $1+3$ covariant formalism}
This analogy can be further worked out by making ``spatial'' and ``temporal''
projections of both the Maxwell equations (\ref{fe3}) and the higher order
gravitational equations (\ref{fe4}). Using the 1+3 covariant formalism (see
e.g. \cite{Maartens:1997fg} and references therein), such
projections can be made and still keep covariance. To explain this formalism,
keeping this paper self-contained, we  need to introduce the notation given in
\cite{Maartens:1997fg}.

Consider a congruence of observers with 4-velocity $U^{\mu}$, and the
\textit{projector into local rest frames} $h_{\mu \nu}\equiv g_{\mu
\nu}+U_{\mu}U_{\nu}$. Note that
\bequ
h_{\mu \nu}U^{\mu}=0 \ , \ \ \ \ \ h_{\mu}^{\ \nu}h_{\nu}^{\
\alpha}=h_{\mu}^{\
\alpha} \ , \ \ \ \ \ h_{\mu \nu}g^{\nu \alpha}=h_{\mu}^{\ \alpha} \ . \eequ
The fundamental idea is that $U^{\alpha}$ provides a covariant time projection
whereas $h_{\alpha \beta}$ provides a covariant spatial projection of
tensorial
quantities. Thus, denote the spatially projected part of a vector as
$V_{\langle
\mu \rangle}\equiv h_{\mu}^{\ \nu}V_{\nu}$ and the spatially projected,
symmetric and trace free part of a rank two tensor as
\bequ
A_{\langle \mu \nu\rangle}\equiv h_{(\mu}^{\ \ \alpha}h_{\nu)}^{\
\beta}A_{\alpha \beta}-\frac{1}{3}h_{\mu \nu}h_{\alpha \beta}A^{\alpha
\beta} \
. \eequ
The covariant spatial vector product is denoted
\bequ
[V,W]_{\mu}\equiv \epsilon_{\mu\nu\alpha\beta}V^{\nu}W^{\alpha}U^{\beta} \ ,
\eequ
where $\epsilon_{\mu\nu\alpha\beta}$ is the Levi-Civita tensor. The covariant
vector product for spatial tensors is
\bequ
[A,B]_{\mu}\equiv \epsilon_{\mu\nu\alpha\beta}A^{\nu}_{\
\sigma}B^{\alpha\sigma}U^{\beta} \ . \eequ
The covariant time derivative for an arbitrary tensor is
\bequ
\dot{A}^{\mu\dots}_{\ \ \ \nu
\dots}=U^{\alpha}\nabla_{\alpha}A^{\mu\dots}_{\ \
\ \nu \dots} \ , \eequ
where $\nabla$ denotes (in this
section) the standard spacetime covariant derivative. The covariant spatial
derivative is
\bequ
D_{\alpha}{A}^{\mu\dots}_{\ \ \ \nu \dots}=h_{\alpha}^{\ \sigma}h^{\mu}_{\
\tau}\dots h_{\nu}^{\ \rho}\dots\nabla_{\sigma}A^{\tau\dots}_{\ \ \
\rho \dots}
\ . \eequ
The covariant spatial divergence and curl of vectors is
\bequ
div V=D^{\mu}V_{\mu} \ , \ \ \ \ \ \ \ \ curl V_{\mu}=\epsilon_{\mu
\nu\alpha\beta}D^{\nu}V^{\alpha}U^{\beta} \ , \eequ
whereas the spatial divergence and curl of rank two tensors is
\bequ
(div A)_{\alpha}=D^{\mu}A_{\alpha \mu} \ , \ \ \ \ \ \ \ \ curl A_{\alpha
\beta}=-\epsilon_{\mu \nu\gamma (\alpha}D^{\mu}A_{\beta)}^{\ \
\nu}U^{\gamma} \
. \eequ
Finally, the kinematics of the $U^{\mu}$-congruence is described by the
following quantities: \textit{expansion}, $\Theta\equiv D^{\mu}U_{\mu}$;
\textit{shear}, $\sigma_{\mu \nu}\equiv D_{\langle \mu}U_{\nu\rangle}$;
\textit{vorticity}, $w_{\mu}\equiv -curl U_{\mu}/2$; \textit{acceleration},
$\dot{U}_{\mu}\equiv \dot{U}_{\langle \mu \rangle}$.

We can now express the Maxwell equations in this formalism. The first equation
in (\ref{fe3}) yields, as its covariant time and space components,
respectively
\bequ
\barr{c}
\displaystyle{D_{\mu}E^{\mu}=\rho-2\omega_{\mu}B^{\mu} \ , } \spa{0.4cm}\\
\displaystyle{\dot{E}_{\langle
\mu\rangle}-{\rm{curl}}B_{\mu}=-\frac{2}{3}E_{\mu}\Theta+\sigma_{\mu\nu}E^{\nu}-[\omega,E]_{\mu}+[\dot{U},B]_{\mu}-j_{\mu}
\ , } \earr \label{cm1}\eequ
where the charge density is given by $\rho=-J_{\mu}U^{\mu}$ and the current
density vector is $j_{\mu}=J_{\alpha}h^{\alpha}_{\ \mu}$. The Bianchi
identities in (\ref{fe3}), yield, as their covariant time and space
components,
respectively
\bequ
\barr{c}
\displaystyle{D_{\mu}B^{\mu}=2\omega_{\mu}E^{\mu} \ , } \spa{0.4cm}\\
\displaystyle{\dot{B}_{\langle\mu\rangle}+{\rm{curl}}E_{\mu}=-\frac{2}{3}B_{\mu}\Theta+\sigma_{\mu\nu}B^{\nu}-[\omega,B]_{\mu}-[\dot{U},E]_{\mu}
\ . } \earr \label{cm2}\eequ
Note that that the vector equations in (\ref{cm1}) and
(\ref{cm2}) are spatial and therefore there are 3+1+3+1=8 equations in
accordance to Maxwell's theory.

Now we turn to the gravitational equations (\ref{fe4}) taking the energy
momentum tensor of a perfect fluid
$T_{\mu\nu}=pg_{\mu\nu}+(p+\rho)U_{\mu}U_{\nu}$. Taking the trace of the first
equation in (\ref{fe4}) implies energy-momentum conservation. The trace free
part gives, taking spacial and temporal projections, a set of equations
\cite{Maartens:1997fg} which resemble (\ref{cm1}) and (\ref{cm2}):
\bequ
\barr{c}
\displaystyle{D^{\mu}\mathcal{E}_{\nu\mu}=\frac{1}{3}D_{\nu}\rho-3\omega^{\mu}\mathcal{H}_{\nu\mu}+[\sigma,\mathcal{H}]_{\nu}
\ , } \spa{0.4cm}\\
\displaystyle{\dot{\mathcal{E}}_{\langle
\mu\nu\rangle}-{\rm{curl}}\mathcal{H}_{\mu\nu}=-\mathcal{E}_{\mu\nu}\Theta+3\sigma_{\tau\langle\mu}\mathcal{E}_{\nu\rangle}^{\
\tau}-\omega^{\tau}\epsilon_{\tau\rho(\mu}\mathcal{E}_{\nu)}^{\
\rho}+2\dot{U}^{\rho}\epsilon_{\rho\tau(\mu}\mathcal{H}_{\nu)}^{\
\tau}-\frac{1}{2}(\rho+p)\sigma_{\mu\nu}\ , } \earr \label{gcm1}\eequ
where $\epsilon_{\mu \nu\rho}\equiv \epsilon_{\mu \nu \rho \tau}U^{\tau}$ and
\bequ
\barr{c}
\displaystyle{D^{\mu}\mathcal{H}_{\nu\mu}=(\rho+p)\omega_{\nu}+3\omega^{\mu}\mathcal{E}_{\nu\mu}-[\sigma,\mathcal{E}]_{\nu}
\ , } \spa{0.4cm}\\
\displaystyle{\dot{\mathcal{H}}_{\langle
\mu\nu\rangle}+{\rm{curl}}\mathcal{E}_{\mu\nu}=-\mathcal{H}_{\mu\nu}\Theta+3\sigma_{\tau\langle\mu}\mathcal{H}_{\nu\rangle}^{\
\tau}-\omega^{\tau}\epsilon_{\tau\rho(\mu}\mathcal{H}_{\nu)}^{\
\rho}-2\dot{U}^{\rho}\epsilon_{\rho\tau(\mu}\mathcal{E}_{\nu)}^{\ \tau}\ . }
\earr \label{gcm2}\eequ
Equations (\ref{cm1})-(\ref{cm2}) and (\ref{gcm1})-(\ref{gcm2}) exhibit
a clear
analogy. The obvious question is the physical
content of this analogy.

\subsection{What physical conclusions can we extract from this analogy?}
\label{What physical conclusions}

The physical content of the analogy
$\{E^{\alpha},\,B^{\alpha}\}\leftrightarrow\{\mathcal{E}^{\mu \nu},\,
\mathcal{H}^{\mu \nu}\}$ is an unanswered question in the literature, and that
is mainly due to the fact that (especially) the magnetic part of the Weyl
tensor is not well understood
\cite{LozanovskiAarons:99,Dunsby:1998hd,VandenBergh:2002fb}.
On some of the literature where this analogy is pursued, it has been suggested
(see e.g.
\cite{Bonnor:95,Lozanovski:99,Maartens:1997fg,Fodor:1998jp,Matarrese:1993zf})
that rotation should source the
magnetic part of the Weyl tensor; but immediately contradictions arise.
Whereas a number of examples, like the Kerr metric or the Van-Stockum solution
\cite{Bonnor:95}, are known to support the idea that rotation sources
$\mathcal{H}_{\mu \nu}$, there are also well known counterexamples, like the
G\"odel universe \cite{Bonnor:95,Lozanovski:99,Maartens:1997fg}. Should we be
surprised?

In the framework of our analogy based on tidal tensors, both the
questions of 1)
the physical interpretation of $\mathcal{E}_{\mu \nu}$ and $\mathcal{H}_{\mu
\nu}$ and 2) the physical content of this analogy based on the
splitting of the
Weyl tensor are readily answered.
\begin{enumerate}
\item $\mathcal{E}_{\mu \nu}$ and $\mathcal{H}_{\mu \nu}$ are tidal
tensors. As
equations (\ref{ERiemann-EWeyl})-(\ref{HRiemann-HWeyl}) point out, they are,
respectively, the trace-free and symmetric parts of $\mathbb{E}_{\alpha\beta}$
and $\mathbb{H}_{\alpha\beta}$, which, in their turn, are the gravitational
counterparts of the electromagnetic tidal tensors $E_{\mu\nu}$ and
$B_{\mu\nu}$.
\item It follows that the analogy
$\{E^{\alpha},\,B^{\alpha}\}\leftrightarrow\{\mathcal{E}^{\mu \nu},\,
\mathcal{H}^{\mu \nu}\}$ is purely formal, since it compares electromagnetic
\textit{fields} with gravitational \textit{tidal} tensors.
\end{enumerate}

 From 1) it is then quite clear why $\mathcal{H}_{\mu \nu}$ is not
zero for Kerr (which is rotating) but should vanish for G\"odel, which is also
interpreted as being rotating. The electromagnetic analogue for the
(asymptotic)
Kerr metric is a rotating
charge (see \ref{rotatingp}), whose magnetic tidal tensor is not zero. But the
electromagnetic
analogue for G\"odel is a uniform magnetic field (see section \ref{Godel}),
whose magnetic tidal tensor
$\mathbb{H}_{\alpha\beta}$ is obviously vanishing; this is the reason for the
vanishing of the magnetic part of the Weyl tensor since, by virtue of equation
(\ref{HRiemann-HWeyl}),
$\mathcal{H}_{\alpha\beta}=\mathbb{H}_{(\alpha\beta)}$.
One can also check that
$\mathcal{H}_{\mu \nu}=0$ for the Heisenberg spacetime, which is again
analogous to a uniform magnetic field.

Evidence for 2) is the simple observation that the gravitational
equation analogous to Gauss's law (\ref{cm1}), $\nabla\cdot\vec{E}=
\rho+\dots$
is (\ref{gcm1}), $D^{\mu}\mathcal{E}_{\nu\mu}=\frac{1}{3}D_{\nu}\rho+\dots$,
clearly revealing that these gravitational objects are one order higher in
differentiation than their electromagnetic counterparts. Another signature of
the unphysical character of this analogy is that the invariants (\ref{inv3})
are not similar, in simple gravitational backgrounds with an
obvious electromagnetic analogue, to the invariants (\ref{inv}) of the
electromagnetic
analogue (see section \ref{invk}).

Under debate\footnote{To follow this debate, see \cite{general relativity and
cosmology} pp. 124-135,
\cite{Bertschinger:1994nc}-\cite{Ellis:Newtonian Weyl},
\cite{Dunsby:1998hd}, and \cite{Maartens:1998ci}.} in literature has been the
question of whether it is possible to neglect $\mathcal{H}_{\mu\nu}$ in the
Newtonian limit; we believe our interpretation of $\mathcal{H}_{\mu \nu}$ may
also shed light into this debate: $\mathcal{H}_{\mu \nu}$ is a magnetic tidal
tensor, thus, motion of mass/energy is its source; Newtonian gravity, like
electrostatics, is obtained by neglecting all relativistic effects; that
includes magnetism\footnote{see \cite{Schutz} p. 246 and \cite{Feynman} 13-6
for insightful discussions on how magnetism arises from relativistic
effects in
gravity and in electromagnetism.}.  Hence, from the viewpoint of our analogy
based on tidal tensors, $\mathcal{H}_{\mu \nu}$ must vanish in the Newtonian
limit for the same reason that $B_{\alpha\beta}$ (or the magnetic field
$\vec{B}$) has no place in electrostatics.
This supports what is asserted by most authors (see
\cite{LozanovskiAarons:99}-\cite{Matarrese:1992rp}
\cite{Bruni:1994nf} \cite{Maartens:1996uv}-\cite{VandenBergh:2002fb}
\cite{Ellis:Newtonian Weyl,Ferrando:03}), and the claim by Ellis and Dunsby
\cite{Ellis:Newtonian Weyl} that the ``Newtonian'' equations for
$\mathcal{H}_{\mu \nu}$ derived in \cite{Bertschinger:1994nc,Hui:1996} were in
fact obtained by going beyond Newtonian theory.

\section{Invariants and a new proof of a set of tensor identities}
\label{secinv}

The fact that both the contractions of $E^{\alpha}$ and $B^{\alpha}$ given in
(\ref{inv}) and the contractions of $\mathcal{E}^{\alpha\beta}$ and
$\mathcal{H}^{\alpha\beta}$ given in (\ref{inv3}) form $U^{\mu}$ independent
tensor quantities, suggests that there
might be some general underlying structure common to both. To see that this is
indeed the case we prove the following lemma:

\bigskip
\begin{description}
\item[{\bf Lemma:}]
Let $A_{\alpha \beta}$ and $B_{\alpha \beta}$ be tensors which are
antisymmetric
in two of their indices; these are the only indices displayed. Let $\star
A_{\alpha \beta}$ and $\star B_{\alpha \beta}$ be the four dimensional Hodge
duals of $A$ and $B$ with respect to these two indices, i.e
\bequ
\star A^{\alpha \beta}=\frac{1}{2}\epsilon^{\alpha \beta \mu \nu}A_{\mu
\nu} \ ,
\eequ
and similarly for $B$. Then, in four dimensions
\bequ
\star A^{\alpha \nu}B_{\alpha \mu}+\star B^{\alpha \nu}A_{\alpha
\mu}=\frac{1}{4}\left[\star A^{\sigma \tau}B_{\sigma \tau}+\star B^{\sigma
\tau}A_{\sigma \tau}\right]\delta_{\mu}^{\nu} \ . \eequ

\bigskip

\item[{\bf Proof:}]
Take the following four dimensional identity:
\bequ
\epsilon^{\alpha \beta \sigma \tau}A_{[\alpha \beta}B_{\sigma
\tau}\delta_{\mu]}^{\nu} \stackrel{D=4}{\equiv} 0 \ . \eequ
The anti-symmetrisation in five indices guarantees the identity in four
dimensions; $\epsilon$ is the Levi-Civita tensor (not the tensor density). A
tedious, but straightforward, computation shows that
\bequ
\epsilon^{\alpha \beta \sigma \tau}A_{[\alpha \beta}B_{\sigma
\tau}\delta_{\mu]}^{\nu}=\frac{6}{30}\left(\left[\star A^{\sigma
\tau}B_{\sigma
\tau}+\star B^{\sigma \tau}A_{\sigma \tau}\right]\delta_{\mu}^{\nu}-4\star
A^{\alpha \nu}B_{\alpha \mu}-4\star B^{\alpha \nu}A_{\alpha \mu}\right) \ ,
\eequ
which proves the Lemma.
\end{description}

We can now apply this lemma to different tensors $A_{\alpha \beta}$ and
$B_{\alpha \beta}$. In doing so we will understand why the contractions
(\ref{inv}) and (\ref{inv3}) are $U^{\alpha}$ independent, as well as build
others.

\begin{description}
\item[{\bf Corollary 1:}] Take $A_{\alpha \beta}=B_{\alpha \beta}=F_{\alpha
\beta}$, the Maxwell tensor. One gets the identity
\bequ
\star F^{\sigma \nu}F_{\sigma \mu}=\frac{\star F^{\sigma \tau}F_{\sigma
\tau}}{4}\delta_{\mu}^{\nu} \ . \label{cor1}\eequ

\item[{\bf Corollary 2:}] Take $A_{\alpha \beta}=F_{\alpha \beta}$ and
$B_{\alpha \beta}=\star F_{\alpha \beta}$. One gets the identity
\bequ
F^{\sigma \nu}F_{\sigma \mu}-\star F^{\sigma \nu}\star F_{\sigma
\mu}=\frac{F^{\sigma \tau}F_{\sigma \tau}}{2}\delta_{\mu}^{\nu} \
.\label{cor2}
\eequ
\end{description}
Contracting (\ref{cor1}) and (\ref{cor2}) with a normalised four velocities
$U^{\mu}$ and $U_{\nu}$ one finds (\ref{inv}). In particular, one
realises that
the important point in getting a scalar invariant which is independent on the
four velocity is the antisymmetric structure of the tensors $A_{\alpha \beta}$
and $B_{\alpha \beta}$ as well as the normalisation of the four velocity.
Likewise we can understand the construction of the invariants
(\ref{inv3}), but
now one uses a two step process:

\begin{description}
\item[{\bf Corollary 3a:}] Take $A_{\alpha \beta}=C_{\alpha \beta
\gamma \eta}$
and $B_{\alpha \beta}=C_{\alpha \beta}^{\ \ \ \tilde{\gamma} \tilde{\eta}}$,
the Weyl tensor. Contracting $\gamma$ with $\tilde{\gamma}$ and $\eta$ with
$\tilde{\eta}$ one gets the identity
\bequ
\star C^{\sigma \nu \gamma \eta}C_{\sigma \mu \gamma \eta}=\frac{\star
C^{\sigma
\tau \gamma \eta}C_{\sigma \tau \gamma \eta}}{4}\delta_{\mu}^{\nu} \ .
\label{weyl1} \eequ

\item[{\bf Corollary 3b:}] Take $A_{\alpha \beta}=C_{\alpha \beta
\gamma \eta}$
and $B_{\alpha \beta}=C_{\alpha \beta}^{\ \ \ \tilde{\gamma} \tilde{\eta}}$.
Contracting $\gamma$ with $\tilde{\gamma}$ (but not $\eta$ with
$\tilde{\eta}$)
one gets the identity
\bequ
\star C^{\sigma \nu}_{\ \ \ \gamma \eta}C_{\sigma \mu}^{\ \ \ \gamma
\tilde{\eta}}+\star C^{\sigma \nu \gamma \tilde{\eta}}C_{\sigma \mu \gamma
\eta}=\frac{1}{4}\left[\star C^{\sigma \tau}_{\ \ \ \gamma \eta}C_{\sigma
\tau}^{\ \ \ \gamma \tilde{\eta}}+\star C^{\sigma \tau \gamma
\tilde{\eta}}C_{\sigma \tau \gamma \eta}\right]\delta_{\mu}^{\nu} \ . \eequ
Using (\ref{weyl1}), one obtains the identity
\bequ
\star C^{\sigma \nu}_{\ \ \ \gamma \eta}C_{\sigma \mu}^{\ \ \ \gamma
\tilde{\eta}}+\star C^{\sigma \nu \gamma \tilde{\eta}}C_{\sigma \mu \gamma
\eta}=\frac{\star C^{\sigma \tau \gamma \eta}C_{\sigma \tau \gamma
\eta}}{8}\delta_{\mu}^{\nu}\delta_{\eta}^{\tilde{\eta}} \ . \label{cor3}\eequ

\item[{\bf Corollary 4a:}] Take $A_{\alpha \beta}=C_{\alpha \beta
\gamma \eta}$
and $B_{\alpha \beta}=\star C_{\alpha \beta}^{\ \ \ \tilde{\gamma}
\tilde{\eta}}$. Contracting $\gamma$ with $\tilde{\gamma}$ and $\eta$ with
$\tilde{\eta}$ one gets the identity
\bequ
C^{\sigma \nu \gamma \eta}C_{\sigma \mu \gamma \eta}-\star C^{\sigma
\nu \gamma
\eta}(\star C)_{\sigma \mu \gamma \eta}=\frac{C^{\sigma \tau \gamma
\eta}C_{\sigma \tau \gamma \eta}}{2}\delta_{\mu}^{\nu} \ . \label{weyl2} \eequ

\item[{\bf Corollary 4b:}] Take $A_{\alpha \beta}=C_{\alpha \beta
\gamma \eta}$
and $B_{\alpha \beta}=\star C_{\alpha \beta}^{\ \ \ \tilde{\gamma}
\tilde{\eta}}$. Contracting $\gamma$ with $\tilde{\gamma}$ (but not
$\eta$ with
$\tilde{\eta}$) one gets the identity
\bequ
\star C^{\sigma \nu}_{\ \ \ \gamma \eta}\star C_{\sigma \mu}^{\ \ \ \gamma
\tilde{\eta}}-C^{\sigma \nu \gamma \tilde{\eta}}C_{\sigma \mu \gamma
\eta}=\frac{1}{4}\left[\left(\star C\right)^{\sigma \tau}_{\ \ \ \gamma
\eta}\star C_{\sigma \tau}^{\ \ \ \gamma \tilde{\eta}}-C^{\sigma \tau \gamma
\tilde{\eta}}C_{\sigma \tau \gamma \eta}\right]\delta_{\mu}^{\nu} \ . \eequ
Using (\ref{weyl2}), one obtains the identity
\bequ
C^{\sigma \nu \gamma \tilde{\eta}}C_{\sigma \mu \gamma \eta}-
\star C^{\sigma \nu}_{\ \ \ \gamma \eta}\star C_{\sigma \mu}^{\ \ \ \gamma
\tilde{\eta}}=\frac{C^{\sigma \tau \gamma \eta}C_{\sigma \tau \gamma
\eta}}{8}\delta_{\mu}^{\nu}\delta_{\eta}^{\tilde{\eta}} \ . \label{cor4}\eequ
\end{description}
The invariants (\ref{inv3}) follow, now, by contracting (\ref{cor3}) and
(\ref{cor4}) with normalised four velocities $U^{\mu}$, $U_{\nu}$, $U^{\eta}$
and $U_{\tilde{\eta}}$. Again, the important underlying structure is the
antisymmetry and the normalisation of the four velocities.

Having understood the structure in the construction of the above
invariants, we
can in an obvious fashion build another two invariants:

\begin{description}
\item[{\bf Corollary 5:}] Take $A_{\alpha \beta}=F_{\alpha \beta; \eta}$ and
$B_{\alpha \beta}=F_{\alpha \beta}^{\ \ \  ; \tilde{\eta}}$.
Contracting $\eta$
with $\tilde{\eta}$ one gets the identity
\bequ
\star F^{\sigma \nu;\eta}F_{\sigma \mu;\eta}=\frac{\star F^{\sigma
\tau;\eta}F_{\sigma \tau;\eta}}{4}\delta_{\mu}^{\nu} \ .\label{cor5} \eequ

\item[{\bf Corollary 6:}] Take $A_{\alpha \beta}=F_{\alpha \beta;\eta}$ and
$B_{\alpha \beta}=\star F_{\alpha \beta}^{\ \ \ ;\tilde{\eta}}$.  Contracting
$\eta$ with $\tilde{\eta}$ one gets the identity
\bequ
F^{\sigma \nu;\eta}F_{\sigma \mu;\eta}-\star F^{\sigma \nu;\eta}\star
F_{\sigma
\mu;\eta}=\frac{F^{\sigma \tau;\eta}F_{\sigma \tau;\eta}}{2}\delta_{\mu}^{\nu}
\ . \label{cor6}\eequ

\end{description}
Contracting with the normalised four velocities $U^{\mu}$, $U_{\nu}$ we
find the
invariants:

\begin{description}
\item[$\bullet$] From corollary 5 and 6:
\bequ
L\equiv E_{\alpha \beta}E^{\alpha \beta}- B_{\alpha \beta}B^{\alpha
\beta}=-\frac{F^{\sigma \tau;\eta}F_{\sigma \tau;\eta}}{2} \ , \ \ \ \ \ \ \ \
M\equiv E_{\alpha \beta}B^{\alpha \beta}=-\frac{\star F^{\sigma
\tau;\eta}F_{\sigma \tau;\eta}}{4} \ ; \label{invemgtidal}\eequ
\end{description}
In the same way that the other invariants were:
\begin{description}
\item[$\bullet$] From corollary 1 and 2 the invariants based on the
electromagnetic fields:
\bequ
L_F \equiv E_{\mu}E^{\mu}-B_{\mu}B^{\mu}=-\frac{F^{\sigma \tau}F_{\sigma
\tau}}{2} \ , \ \ \ \ \ \ \ M_F \equiv E_{\mu}B^{\mu}=-\frac{\star F^{\sigma
\tau}F_{\sigma \tau}}{4} \ ; \label{invemg}\eequ
\item[$\bullet$] From corollary 3 and 4:
\bequ
\mathcal{L}\equiv \mathcal{E}_{\alpha \beta}\mathcal{E}^{\alpha \beta}-
\mathcal{H}_{\alpha \beta}\mathcal{H}^{\alpha \beta}=\frac{C^{\sigma \tau
\gamma \eta}C_{\sigma \tau \gamma \eta}}{8} \ , \ \ \ \ \ \ \ \ \
\mathcal{M}\equiv \mathcal{E}_{\alpha \beta}\mathcal{H}^{\alpha
\beta}=\frac{\star C^{\sigma \tau \gamma \eta}C_{\sigma \tau \gamma \eta}}{16}
\ . \label{invweyl}\eequ

\end{description}
All these quantities are independent of the observer $O$ whose four
velocity is
$U^{\alpha}$ despite the fact that the definition of electric and magnetic
tidal tensors (\ref{Eab})-(\ref{Bab}), electric and magnetic
parts of the Maxwell tensor (\ref{split}) and electric and magnetic parts of
the Weyl tensor (\ref{Weyl E H}) depend on $O$.

\subsection{Invariants for the gravito-electromagnetic analogy based on tidal
tensors}
\label{Invariants for our analogy}

The previous section established that the observer independence of the
invariants (\ref{invemg}) and (\ref{invweyl}) steams from the
underlying tensor
structure rather than from some physical similarity between
$\left\{ E_{\alpha},B_{\alpha}\right\}$ and
$\left\{ \mathcal{E}_{\alpha\beta},\mathcal{H}_{\alpha\beta}\right\}$;
in fact,
the example \ref{invk}, below, shows that they do not match.

A natural question is if it is possible to define, in electromagnetism,
invariants physically analogue to $\mathcal{L}$ and $\mathcal{M}$. In vacuum,
$\mathcal{E}_{\alpha\beta}$ and $\mathcal{H}_{\alpha\beta}$ are the
gravitational tidal tensors. Since we have shown in (\ref{invemgtidal}) that
the electromagnetic tidal tensors also define the invariants $L$ and $M$, then
these should be the physical analogues of the
invariants $\mathcal{L}$
and $\mathcal{M}$. The example \ref{invk} supports this claim. Thus, we
establish the analogy:
\begin{eqnarray*}
L\equiv E^{\alpha\beta}E_{\alpha\beta}-E^{\alpha\beta}H_{\alpha\beta} &
\leftrightarrow &
\mathcal{L}\equiv\mathcal{E}^{\alpha\beta}\mathcal{E}_{\alpha\beta}-\mathcal{H}^{\alpha\beta}\mathcal{H}_{\alpha\beta}\\
M\equiv E^{\alpha\beta}H_{\alpha\beta} & \leftrightarrow &
\mathcal{M}\equiv\mathcal{E}^{\alpha\beta}\mathcal{H}_{\alpha\beta}\end{eqnarray*}
which holds in vacuum. One then might guess that, in the presence of sources,
replacing
$\left\{ \mathcal{E}_{\alpha\beta},\mathcal{H}_{\alpha\beta}\right\}$ by
$\left\{ \mathbb{E}_{\alpha\beta},\mathbb{H}_{\alpha\beta}\right\}$ would lead
to a natural
extension of the analogy with the electromagnetic invariants $L$ and $M$. This
turns out \textit{not} to be the case: in gravity it is not possible,
except in
vacuum, to construct scalar invariants using only the electric and magnetic
tidal tensors. As already discussed in section \ref{secanalogy General
Arguments}, equations (\ref{ERiemann-EWeyl})-(\ref{HRiemann-HWeyl}) show that
the sources endow $\mathbb{E}_{\alpha\beta}$ with a trace and
$\mathbb{H}_{\alpha\beta}$ with an anti-symmetric
part; thus these tensors combined
possess 6+8 independent components which is insufficient to encode all the
information in
the Riemann tensor (20 independent components in four dimensions).  A third
spacial, symmetric tensor,
defined by (see \cite{cherubini:02} and \cite{Gravitation} pp. 360-361):
\bequ
\mathbb{F}_{\alpha\gamma}\equiv\star
R\star_{\alpha\beta\gamma\delta}U^{\beta}U^{\delta} \ ,\eequ
where Hodge duality is taken both with respect to the first and second pair of
indices, is needed to account for the remaining 6 components. It is then
straightforward to show that the
invariants formed by the Riemann tensor are:
\bequ
\mathbb{L}=\frac{R^{\sigma \tau \gamma \eta}R_{\sigma \tau \gamma
\eta}}{8}=\frac{\mathbb{E}^{\alpha\gamma}\mathbb{E}_{\alpha\gamma}+\mathbb{F}^{\alpha\gamma}\mathbb{F}_{\alpha\gamma}}{2}-\mathbb{H}^{\alpha\gamma}\mathbb{H}_{\alpha\gamma}
\ , \ \ \ \
\mathbb{M}=\frac{\star R^{\sigma \tau \gamma \eta}R_{\sigma \tau \gamma
\eta}}{16}=
\frac{\mathbb{E}^{\alpha\gamma}\mathbb{H}_{\alpha\gamma}-\mathbb{F}^{\alpha\gamma}\mathbb{H}_{\alpha\gamma}}{2}
\ . \label{invriem}\eequ
Noting that:
\begin{eqnarray*}
\mathbb{E}^{\alpha\gamma}\mathbb{H}_{\alpha\gamma}=\mathcal{E}^{\alpha\gamma}\mathcal{H}_{\alpha\gamma}-\frac{1}{2}\mathcal{H}^{\alpha\gamma}R_{\alpha\gamma}\,\,,\,\,\,\,\,\,\,\,\,\,\mathbb{F}^{\alpha\gamma}\mathbb{H}_{\alpha\gamma}=-\mathcal{E}^{\alpha\gamma}\mathcal{H}_{\alpha\gamma}-\frac{1}{2}\mathcal{H}^{\alpha\gamma}R_{\alpha\gamma}\,\,.\end{eqnarray*}
it follows that $\mathcal{M}=\mathbb{M}$ generically (i.e. not only in
vacuum).
Note also that in vacuum:
\[
R_{\alpha\beta\gamma\delta}=C_{\alpha\beta\gamma\delta}=-*C*_{\alpha\beta\gamma\delta}\Rightarrow\mathbb{F}_{\alpha\gamma}=-\mathbb{E}_{\alpha\gamma}=-\mathcal{E}_{\alpha\gamma}\]
so that (\ref{invriem}) reduces to (\ref{invweyl}), as expected.

It is a curious fact that the most relevant energy conditions can be expressed
in a simple fashion in terms of the three tensors in which the Riemann tensor
is generically decomposed
($\mathbb{E}_{\alpha\beta},\mathbb{H}_{\alpha\beta},\mathbb{F}_{\alpha\beta}$).
Indeed we have seen from (\ref{MaxGrav}i) that $\mathbb{E}_{\
\alpha}^{\alpha}>0$
is the strong energy condition. A simple computation reveals that
$\mathbb{F}_{\
\alpha}^{\alpha}=8\pi T_{\mu \nu}U^{\mu}U^{\nu}$, which shows that
$\mathbb{F}_{\ \alpha}^{\alpha}>0$ is the weak energy
condition. Finally, the dominant energy condition is the statement that the
vector
\[ J^{\nu}=\frac{1}{8\pi}\left(\mathbb{H}_{\alpha\beta}U_{\mu}\epsilon^{\alpha
\beta
\mu \nu}+ \mathbb{F}_{\ \alpha}^{\alpha}U^{\nu}\right)\  \]
is timelike and future directed, where we have used (\ref{MaxGrav}iv).

\subsubsection{A simple example}
\label{invk}
Having exhibited all the above invariants we now return to the example of
section \ref{rotatingp}. In the electromagnetic case, we can compute both the
usual relativistic invariants
\begin{equation}
L_F=\frac{q^{2}}{r^{4}}-\frac{\mu^{2}\left(5+3\cos2\theta\right)}{2r^{6}}\label{E^2-B^2}
\ , \ \ \ \ \ \ M_F=\frac{2\mu q\cos\theta}{r^{5}}\label{E.B} \ ,
\end{equation}
as well as the invariants based on the tidal tensors
\begin{equation}
L=\frac{6q^{2}}{r^{6}}-\frac{\mu^{2}\left(6+3\cos2\theta\right)}{r^{8}}
\ , \ \ \ \ \ \ M=\frac{18\mu q\cos\theta}{r^{7}} \ . \end{equation}
On the gravitational side, since there are no sources, we have the asymptotic
expressions:
\begin{equation}
\mathcal{L}=\mathbb{L} \approx\frac{6m^{2}}{r^{6}} \ , \  \ \ \ \ \ \
\mathcal{M}=\mathbb{M}\approx\frac{18Jm\cos\theta}{r^{7}}\label{MKerr} \ .
\end{equation}
Identifying the black hole mass $m$ and angular momentum $J$ with,
respectively,
the electromagnetic charge $q$ and the magnetic dipole moment $\mu$, the
asymptotic matching between $\mathcal{L}$ and $L$ (not $L_F$!), and between
$\mathcal{M}$ and $M$ (not $M_F$!), is perfect. It is a  curious fact
that the matching of the invariants is \textit{exact} between
Schwarzschild and
a point charge; the non-vanishing invariants are
\bequ
L=\frac{6q^{2}}{r^{6}} \ , \ \ \ \ \mathcal{L}=\mathbb{L}=\frac{6m^{2}}{r^{6}}
\ . \eequ
By contrast $L_F=q^{2}/r^{4}$. Thus, it is clear from this example that the
\textit{physical} analogues of the gravitational invariants
$\{\mathcal{L},\mathcal{M}\}$ (in vacuum) are the invariants $\{L,M\}$ built
from tidal tensors, and \textit{not} the invariants $\{L_F,M_F\}$ built from
electromagnetic fields.

\section{Discussion}
In this paper we have proposed a tensorial description of the \emph{physical}
gravitoelectromagnetism (GEM) - a new analogy between general relativity
and electromagnetism, based on tidal tensors. \emph{Physical}
gravitoelectromagnetism
has been studied in literature in the framework of linearised gravity (section \ref{linearth}),
where the analogy $\{ \vec{E},\,\vec{B}\} \leftrightarrow\{
\vec{E}_{G},\,\vec{B}_{G}\}$
is drawn. However, this analogy is non-covariant, and valid only for
weak \emph{static} fields and stressless sources.
There is another well known analogy between gravity and electromagnetism: $\left\{
E^{\alpha},\, B^{\alpha}\right\} \leftrightarrow\left\{
\mathcal{E}^{\mu\nu},\,\mathcal{H}^{\mu\nu}\right\} $, which
has also been dubbed gravitoelectromagnetism (section \ref{secanalogy2});
this analogy is covariant and general, but (as we have discussed in
section \ref{What physical conclusions}) it is purely \emph{formal}.
Hence, there is a void in literature concerning a general, physical,
gravito-electromagnetic analogy. Our proposal, the analogy $\left\{
E^{\alpha\beta},\, B^{\alpha\beta}\right\} \leftrightarrow\left\{
\mathbb{E}^{\alpha\beta},\,\mathbb{H}^{\alpha\beta}\right\} $,
fills that void, since it is a \emph{physical} analogy, and relies
on \emph{covariant}, \emph{universal} equations.

The analogy based in tidal tensors naturally embodies all the correct
predictions from the linearised
theory approach, but, since it is exact and general, allows for an
understanding of gravitomagnetism beyond the limits of the latter. The
Som Raychaudhuri/Van-Stockum and G\"odel solutions are examples of universes
which are beyond the scope of the linear approach, and that we were
able to study under our approach (sections \ref{Van-Stockum} and
\ref{Godel}). And even in the linear limit, if the configurations
are time dependent, one must use the analogy based on tidal tensors,
and not the analogy $\{ \vec{E},\,\vec{B}\} \leftrightarrow\{
\vec{E}_{G},\,\vec{B}_{G}\} $,
to obtain correct results (see section \ref{Maxwellian gravity}).

We have shown that the Maxwell equations can be regarded as tidal
tensor equations, and, in this framework, we have derived their gravitational
analogues (section \ref{Gravitational and Electromagnetic Tidal Tensors}). It
follows that the (non-covariant) equations derived form
the analogy $\{ \vec{E},\,\vec{B}\} \leftrightarrow\{
\vec{E}_{G},\,\vec{B}_{G}\} $
turn out to be a special case of our \emph{exact} equations in the
regime of static, weak fields, and stressless sources - which is, therefore,
the regime of validity of the latter approach. This regime of validity
is hence revealed in an unambiguous way, thus shedding light into an
ongoing debate (cf. section \ref{linearth}).

In the framework of this analogy, suggestive similarities between
gravity and electromagnetism have been revealed:

\begin{itemize}
\item For stationary configurations (in the observer rest frame), the
gravitational
and electromagnetic tidal tensors obey strikingly similar equations
(section \ref{Gravity vs Electromagnetism});
\item In vacuum, the gravitational tidal tensors form invariants in the
same way the electromagnetic ones do (section \ref{Invariants for our
analogy}).
\item As an illustration of the previous points, the gravitational tidal
tensors of a Kerr black hole asymptotically match the electromagnetic
tidal tensors from a spinning charge (identifying the appropriate
parameters, cf. section \ref{rotatingp}). And there is also a matching
between observer independent quantities: the invariants built on those
tidal tensors (section \ref{invk}).
\item The Klein Gordon equation reduces, in ultra-stationary spacetimes,
to the non-relativistic Schr\"odinger equation for a particle subject to a
certain magnetic field, living in a curved three-space; the tidal
tensor of that magnetic field turns out to match exactly (up to the
usual factor of 2) the magnetic part of the Riemann tensor (section
\ref{ultra}).
\end{itemize}
But our approach also unveiled deep differences between the two interactions.
These differences are revealed, in a very clear fashion, by the symmetries
of the tidal tensors. Unlike its electromagnetic analogue, the gravitational
electric tidal tensor is always symmetric; in vacuum, the same applies
to the gravitational magnetic tidal tensor. This means that in gravity
there can be no electromagnetic-like induction effects: the
``gravito-electromagnetic''
induced fields which have been predicted in the literature (see section
\ref{Maxwellian gravity}), and whose
detection was recently experimentally attempted, do not exist.

One should not be surprised: the fact that in generic dynamics gravity
cannot be described by electromagnetic like fields, just reminds us
that since gravity is pure geometry, such fields or forces have no
place in it. This does not preclude, of course, a unification of these
two interactions, as in Kaluza-Klein theory; but it gives us a strong
hint that a possible geometrisation of electromagnetism (at least
in four dimensions) would have to take a very different character
from the geometrisation of gravity.

Nevertheless, despite the intrinsic differences between the two interactions,
a physical gravito-electromagnetic analogy is still of great value
for the understanding of both theories. The non-geodesic motion of
a gyroscope is an example of an effect which we showed in section
\ref{Stern-Gerlach} that can not only be easily understood,
but also exactly described, in an analogy with the more familiar
electromagnetic
force exerted on a magnetic dipole. The latter is generated by
non-homogeneities
in the electromagnetic field - it is therefore a purely tidal effect, and
hence the most obvious application for our tidal tensor based analogy.
Indeed, a simple application of our analogy leads to the Papapetrou
equation for the force applied on a gyroscope. There have been previous
attempts to describe this force in an analogy with electromagnetism;
a first order estimate has been derived in the framework of the standard
linearised theory approach to GEM (cf. section \ref{standard approach});
that expression, however, is valid only when the gyroscope is at \emph{rest}
in a static, weak gravitational field (by contrast, our result is
exact, thus valid for fields arbitrary large, and with an arbitrary
time dependence) and therefore not suited
to describe motion. Moreover that force accounts only for the coupling
between the intrinsic spin of the source and the spin of the gyroscope
(that is why the force on a gyroscope, and the violation of the weak
equivalence principle, are often referred in the literature as arising
from a spin-spin interaction) hiding the fact that the gyroscope will
indeed deviate from geodesic motion even in the absence of rotating
sources; for example, in the Schwarzschild spacetime. The underlying
reason is readily understood in the framework of the approach based
on tidal tensors. We derived two fundamental results, (\ref{ForceGR})
and (\ref{ForceEM}), which have a very important physical interpretation:
it is the magnetic tidal tensor as seen by the dipole/gyroscope what
completely determines the force exerted upon it. Therefore, the gyroscope
deviates from geodesic motion in Schwarzschild spacetime by the very
same reason that a magnetic dipole will suffer a Stern-Gerlach type deviation
even in the coulomb field of
a point charge: in its rest frame, there is a non-vanishing magnetic tidal
tensor.

Another important physical content unveiled by the analogy based
on tidal tensors (and which is, again, lost in the three-dimensional expression
derived the linear theory approach) concerns the temporal component
of these forces. In the dipole rest frame, the time component of
(\ref{ForceEM})
is the power transferred to the dipole by Faraday's induction, and
the fact that it is zero in the gravitational case (\ref{ForceGR}),
may be regarded as another evidence for the absence of electromagnetic-like
induction effects in gravity.

Both this example of the force exerted on a dipole/gyroscope, and
the worldline deviation equations (\ref{desvio geo})-(\ref{desvio EM})
exhibit one of the strongest aspects of our analogy: the electromagnetic
and gravitational tidal tensors always play analogous roles in dynamics,
despite being in general very different (they do not even exhibit
generically the same symmetries).

There are many other effects which can be easily assimilated with
the help of a physical gravito-electromagnetic analogy (indeed, the
variety is far too wide to be herein discussed at length). Amongst
the best known ones are the \textit{dragging of inertial frames} and
\textit{gyroscope precession} in the vicinity of rotating bodies,
by analogy with Larmor orbits of charged particles and precession
of magnetic dipole in magnetic fields (section \ref{minper}). Another example,
explored in
this paper, is the G\"odel universe (section \ref{Godel}). It is
commonly found
in the literature
that the G\"odel universe should be interpreted as a rotating non-expanding
universe. Since it is homogeneous one is inevitably led to the conclusion
that it must be rotating around \textit{every point}! Herein, we have
suggested that a more insightful interpretation is as follows. The
magnetic gravitational tidal tensor is zero for the G\"odel Universe.
Thus, one
may think on the G\"odel Universe
as a gravitational version of a constant magnetic field (in a curved
space). This gives, for the rotation of test particles around any
point, a more intuitive picture than the aforementioned
one. As a final example consider the
similarity between the Coulomb and the Newtonian gravitational potential
and also the analogy between the Biot-Savart law and the force between
two mass currents. Note that in both cases the electromagnetic force
(for charges of equal sign) has the \textit{opposite} sign to its
gravitational analog. Thus, two parallel mass currents will attract
one another if they move with opposite velocity. This
builds a physical intuition to explain why, in rotating black holes,
test particles with \textit{counter rotating angular momentum can
generically reach closer to the black hole} (see \cite{Gibbons:1999uv}
for an explicit discussion of this point).

The approach proposed herein clarifies several issues concerning the
gravito-electromagnetic approaches found in the literature. As mentioned
above, it reveals in an unambiguous way the range of validity of the
analogy $\{ \vec{E},\,\vec{B}\} \leftrightarrow\{
\vec{E}_{G},\,\vec{B}_{G}\} $,
for which there is no consensus in the literature. It also sheds light
on some conceptual difficulties concerning the analogy $\left\{ E^{\alpha},\,
B^{\alpha}\right\} \leftrightarrow\left\{
\mathcal{E}^{\mu\nu},\,\mathcal{H}^{\mu\nu}\right\} $,
by giving simple answers to three longstanding questions debated in
the literature (see section \ref{What physical conclusions}), namely:
1) the source of the magnetic part of the Weyl tensor, 2) its vanishing
in homogeneous rotating universes and 3) its Newtonian limit: 1) it
is a tidal magnetic tensor; thus, motion (of masses or transfer of
momentum) is, generically, its source, 2) it vanishes in such spacetimes
because they are analogous to uniform magnetic fields and 3) it has
no place in Newtonian gravity by the same reason that $B_{\alpha\beta}$
has no place in electrostatics, i.e, both limits are obtained neglecting all
relativistic effects, which includes magnetism and (of course) the magnetic
tidal tensors. In the framework of our approach,
it is obvious that the analogy $\left\{ E^{\alpha},\, B^{\alpha}\right\}
\leftrightarrow\left\{ \mathcal{E}^{\mu\nu},\,\mathcal{H}^{\mu\nu}\right\} $
is purely formal, since it compares electromagnetic fields to gravitational
tidal tensors. In section \ref{secinv} we dissected the common
underlying tensorial structure which allows $\left\{
\mathcal{E}^{\mu\nu},\,\mathcal{H}^{\mu\nu}\right\} $
and $\left\{ E^{\alpha},\, B^{\alpha}\right\} $ to form observer
independent scalars in a formally similar way, and showed that indeed,
by the same mathematical reasons, it is also possible to construct
formally similar invariants from the electromagnetic tidal tensors
$\left\{ E^{\mu\nu},\, B^{\mu\nu}\right\} $; these invariants are (in vacuum)
the \emph{physical}
analogues of the invariants formed by the electric and magnetic parts
of the Weyl tensor $\left\{
\mathcal{E}^{\mu\nu},\,\mathcal{H}^{\mu\nu}\right\}
$.

The relation between the two analogies found in literature is now
clear. Take the example of the Heisenberg spacetime (section
\ref{Van-Stockum}).
According to the linearised theory approach, it has a uniform gravito-magnetic
field $\vec{B}_G$ and a vanishing gravito-electric field $\vec{E}_G$; however,
the analogy $\left\{ E^{\alpha},\, B^{\alpha}\right\} \leftrightarrow\left\{
\mathcal{E}^{\mu\nu},\,\mathcal{H}^{\mu\nu}\right\} $ apparently leads to an
opposite conclusion: the magnetic
part of the Weyl tensor, which is therein taken to be the gravitational
analogue of the magnetic field, vanishes, while its electric counterpart
is non-zero. But there is indeed no contradiction. Firstly, the former
is a physical analogy, while the latter is purely formal. Then, the
two approaches refer (on the gravitational side) to different things:
the first, to (fictitious) gravitational fields; the second to tidal
tensors, which are one order higher in differentiation of the metric
potentials than the former. Then, the vanishing of $\mathcal{H}_{\alpha\beta}$
in the second approach is in accordance with the fact that the field
is uniform in the first approach. $\mathcal{E}_{\alpha\beta}$ does
not vanish, unlike one might expect from the vanishing of $\vec{E}_G$ in the
linear theory approach; again there
is no contradiction: it is simply due to the fact that unlike its
magnetic counterpart (in ultra-stationary spacetimes), the electric
tidal tensor is not linear in the metric potentials.

Our approach has also achieved a unification within gravito-electromagnetism.
Indeed, the analogy based on linearised theory (within its range of
validity) was seen to originate from the same fundamental principle
as the (exact) connection between ultra-stationary spacetimes and magnetic
fields in some curved manifolds: the similarity between tidal tensors.

\section*{Acknowledgements}
We are grateful to C. Cherubini and B. Mashhoon for correspondence. C.H. was
partially supported
by the grant SFRH/BPD/5544/2001. This work (and L.C.) was also supported by
\textit{Funda\c c\~ao Calouste Gulbenkian} through \textit{Programa de Est\'\i
mulo \`a Investiga\c c\~ao} and by the \textit{Funda\c c\~ao para a
Ci\^encia e
a Tecnologia} grants POCTI/FNU/38004/2001 and POCTI/FNU/50161/2003. Centro de
F\'\i sica do Porto is partially funded by FCT through the POCTI programme.

\appendix
%dummy comment inserted by tex2lyx to ensure that this paragraph is not empty

\section{Worldline deviation\label{Worldlinedev}}

\subsection{Gravity (geodesic deviation)}

Consider two particles on infinitesimally close geodesics $x^{\alpha}(\tau)$
and $x_{2}^{\alpha}(\tau)=x^{\alpha}(\tau)+\delta x^{\alpha}(\tau)$.
The geodesic equation $DU^{\alpha}/D\tau=0$ yields, for the first
particle:\begin{equation}
0=\frac{d^{2}x^{\alpha}}{d\tau^{2}}+\Gamma_{\mu\beta}^{\alpha}(\mathbf{x})U^{\mu}U^{\beta}\label{geo1}\end{equation}
while for the second particle, by making a first order Taylor expansion
$\Gamma_{\mu\beta}^{\alpha}(\mathbf{x}+\delta\mathbf{x})\approx\Gamma_{\mu\beta}^{\alpha}(\mathbf{x})+\Gamma_{\mu\beta,\sigma}^{\alpha}(\mathbf{x})\delta
x^{\sigma}$,
we obtain:\begin{equation}
0=\frac{d^{2}x^{\alpha}}{d\tau^{2}}+\frac{d^{2}\delta
x^{\alpha}}{d\tau^{2}}+\left[\Gamma_{\mu\beta}^{\alpha}(\mathbf{x})+\Gamma_{\mu\beta,\sigma}^{\alpha}(\mathbf{x})\delta
x^{\sigma}\right]\left[U^{\mu}U^{\beta}+2U^{\mu}\frac{d\delta
x^{\beta}}{d\tau}+\frac{d\delta x^{\mu}}{d\tau}\frac{d\delta
x^{\beta}}{d\tau}\right]\label{Geo2}\end{equation}
subtracting (\ref{geo1}) to (\ref{Geo2}):\begin{eqnarray}
-\frac{d^{2}\delta x^{\alpha}}{d\tau^{2}} & = &
\Gamma_{\mu\beta}^{\alpha}(\mathbf{x})\left[2U^{\mu}\frac{d\delta
x^{\beta}}{d\tau}+\frac{d\delta x^{\mu}}{d\tau}\frac{d\delta
x^{\beta}}{d\tau}\right]\nonumber \\
&  & +\Gamma_{\mu\beta,\sigma}^{\alpha}(\mathbf{x})\delta
x^{\sigma}\left[U^{\mu}U^{\beta}+2U^{\mu}\frac{d\delta
x^{\beta}}{d\tau}+\frac{d\delta x^{\mu}}{d\tau}\frac{d\delta
x^{\beta}}{d\tau}\right]\label{subtract}\end{eqnarray}

The second covariant derivative of the connecting vector along the
geodesic curve $x^{\alpha}(\tau)$ is:\begin{eqnarray*}
\frac{D^{2}\delta x^{\alpha}}{D\tau^{2}} & = & \frac{d^{2}\delta
x^{\alpha}}{d\tau^{2}}-R_{\,\,\,\mu\beta\sigma}^{\alpha}U^{\mu}U^{\sigma}\delta
x^{\beta}+\Gamma_{\sigma\beta,\mu}^{\alpha}U^{\sigma}U^{\beta}\delta
x^{\mu}+2\Gamma_{\mu\beta}^{\alpha}U^{\beta}\frac{d\delta
x^{\mu}}{d\tau}\end{eqnarray*}
Using (\ref{subtract}) we finally get:\begin{eqnarray}
\frac{D^{2}\delta x^{\alpha}}{D\tau^{2}} & = &
-R_{\,\,\,\mu\beta\sigma}^{\alpha}U^{\mu}U^{\sigma}\delta
x^{\beta}-\Gamma_{\mu\beta,\sigma}^{\alpha}\delta
x^{\sigma}\left[2U^{\mu}\frac{d\delta x^{\beta}}{d\tau}+\frac{d\delta
x^{\mu}}{d\tau}\frac{d\delta
x^{\beta}}{d\tau}\right]-\Gamma_{\mu\beta}^{\alpha}\frac{d\delta
x^{\mu}}{d\tau}\frac{d\delta
x^{\beta}}{d\tau}\label{acceleration2}\end{eqnarray}
Although it may not be manifest, this equation is indeed covariant,
as shown in \cite{Generalized Geodesic}. It gives
the {}``relative acceleration'' of two neighbouring geodesics with
arbitrary tangent vectors. If we consider the geodesics to be (at
some instant) parallel, i.e., the two particles to have the same velocity,
then $\frac{d\delta x^{\mu}}{d\tau}=0$ and this equation reduces
to the traditional {}``geodesic deviation equation'':\begin{equation}
\frac{D^{2}\delta
x^{\alpha}}{D\tau^{2}}=-R_{\,\,\,\mu\beta\sigma}^{\alpha}U^{\mu}U^{\sigma}\delta
x^{\beta}\label{acceleration3}\end{equation}

\subsection{Electromagnetism}

We will now consider the analogue electromagnetic problem: two particles
with the same ratio $q/m$ placed in a electromagnetic field on Minkowski
spacetime, following the infinitesimally close worldlines $x^{\alpha}(\tau)$
and $x_{2}^{\alpha}(\tau)=x^{\alpha}(\tau)+\delta x^{\alpha}(\tau)$.
We choose global Cartesian coordinates to perform this analysis. The
equation of motion of the first particle is:\begin{eqnarray}
\frac{d^{2}x^{\alpha}}{d\tau^{2}} & =\frac{q}{m} &
F_{\,\,\,\beta}^{\alpha}(\mathbf{x})U^{\beta}\label{EqMovEM1}\end{eqnarray}

Using a first order Taylor expansion,
$F_{\,\,\,\beta}^{\alpha}(\mathbf{x}+\delta\mathbf{x})\approx
F_{\,\,\,\beta}^{\alpha}(\mathbf{x})+F_{\,\,\,\beta,\sigma}^{\alpha}(\mathbf{x})\delta
x^{\sigma}$;
thus, in this coordinate system, the equation of motion for the second
particle is:\begin{equation}
\frac{d^{2}\left(x^{\alpha}+\delta
x^{\alpha}\right)}{d\tau^{2}}=\frac{q}{m}\left[F_{\,\,\,\beta}^{\alpha}(\mathbf{x})+F_{\,\,\,\beta,\sigma}^{\alpha}(\mathbf{x})\delta
x^{\sigma}\right]\left(U^{\beta}+\frac{d\delta x^{\beta}}{d\tau}\right)\
\label{EqMovEM2}\end{equation}
Subtracting (\ref{EqMovEM1}) to (\ref{EqMovEM2}) we obtain:\begin{equation}
\frac{d^{2}\delta
x^{\alpha}}{d\tau^{2}}=\frac{q}{m}\left[F_{\,\,\,\beta,\sigma}^{\alpha}U^{\beta}\delta
x^{\sigma}+F_{\,\,\,\beta,\sigma}^{\alpha}\delta x^{\sigma}\frac{d\delta
x^{\beta}}{d\tau}+F_{\,\,\,\beta}^{\alpha}\frac{d\delta
x^{\beta}}{d\tau}\right]\label{accelerationEM}\end{equation}
which is the electromagnetic analogue of (\ref{acceleration2}).
Since we are using Cartesian coordinates we can now replace partial
derivatives by covariant derivatives, hence obtaining a manifestly covariant
equation. It gives the acceleration of the vector connecting two
infinitesimally
close particles with arbitrary velocities. If the particles have the
same velocity, then $\frac{d\delta x^{\mu}}{d\tau}=0$ and the deviation
equation reduces to:\begin{equation}
\frac{D^{2}\delta
x^{\alpha}}{D\tau^{2}}=\frac{q}{m}F_{\,\,\,\beta;\sigma}^{\alpha}U^{\beta}\delta
x^{\sigma}\ \label{accelerationEM2}\end{equation}
in a clear analogy with (\ref{acceleration3}).

\end{document}